\documentclass[journal,twocolumn]{IEEEtran}
\usepackage{stfloats,color}
\usepackage{amsfonts}
\usepackage[cmex10]{amsmath}
\usepackage{graphics}
\usepackage{setspace}
\usepackage{cite}
\usepackage{bm}
\usepackage{array}
\usepackage{algorithmic}
\usepackage{mdwmath}
\usepackage{mdwtab}
\usepackage{amssymb}
\usepackage{caption}
\usepackage{subcaption}
\usepackage[font=footnotesize]{subfig}
\usepackage{fixltx2e}
\usepackage{url}
\usepackage{times}
\usepackage{epsfig}
\usepackage{latexsym}
\usepackage{epstopdf}
\usepackage{verbatim}
\usepackage{units}
\usepackage{amsthm}
\usepackage{mdwlist}
\usepackage{tabu}
\usepackage{lipsum,amsmath,multicol}
\usepackage{soul}
\usepackage{flushend}
\usepackage{cuted}
\usepackage{setspace}
\graphicspath{{./figures/}}
\newtheorem*{def:outage}{Definition}
\newtheorem{thm}{Theorem}
\newtheorem*{prof}{Proof}
\newtheorem{lemm}{Lemma}
\begin{document}
\title{Using Network Coding to Achieve the Capacity of Deterministic Relay Networks with Relay Messages}%with Relay Messages}
% <-this % stops a space
%\thanks{Manuscript received April 19, 2005; revised December 27, 2012.}}

\author{Ahmed A.~Zewail,~\IEEEmembership{Student Member,}
        Yahya~Mohasseb,~\IEEEmembership{Senior,}
        Mohammed~Nafie,~\IEEEmembership{Senior,}
        and~Hesham~El Gamal,~\IEEEmembership{Fellow}
  \thanks{A. A. Zewail was with the Wireless Intelligent Networks Center (WINC), Nile University, Giza, Egypt. He is now with the Department of Electrical Engineering, Pennsylvania State University, University Park, PA 16802 USA (email: aiz103@psu.edu). }
\thanks{Y. Mohasseb is with the Department of Communications, The Military Technical College, Cairo, Egypt 11331 (email: mohasseb@ieee.org).}               
\thanks{M. Nafie is with the Wireless Intelligent Networks Center (WINC), Nile University, Giza, Egypt, also he affiliated with the Department of Electronics and Communications, Cairo University (email:mnafie@nileuniversity.edu.eg).   }
\thanks{H. El Gamal is with the Department of Electrical and Computer Engineering, The Ohio State University, Columbus, OH 43210 USA (e-mail: helgamal@ece.osu.edu).}
\thanks{This paper was made possible by NPRP grant $\#$ 4-1119-2-427 from the Qatar National Research Fund (a member of Qatar Foundation). The statement made herein are solely responsibility of the authors.}}
\maketitle

\begin{abstract}
In this paper, we derive the capacity of the deterministic relay networks with relay messages. We consider a network which consists of five nodes, four of which can only communicate via the fifth one. However, the fifth node is not merely a relay as it may exchange private messages with the other network nodes. First, we develop an upper bound on the capacity region based on the notion of a single sided genie. In the course of the achievability proof, we also derive the deterministic capacity of a 4-user relay network (without private messages at the relay). The capacity achieving schemes use a combination of two network coding techniques: the Simple Ordering Scheme (SOS) and Detour Schemes (DS). In the SOS, we order the transmitted bits at each user such that the bi-directional messages
will be received at the same channel level at the relay, while the basic idea behind the DS is that some parts of the message follow an indirect
path to their respective destinations.  This paper, therefore, serves to show that user cooperation and network coding can enhance throughput, even when the users are not directly connected to each other. %, where some parts of the message follow an indirect path to their respective destinations. In the SOS, we order the transmitted bits at each user such that the bi-directional messages will be received at the same channel level at the relay, while the basic idea behind the DS is we sent some bits via alternative paths instead of sending them directly.  
\end{abstract}
% \fi
\IEEEpeerreviewmaketitle
\section{Introduction}
Cooperative communications for wireless networks have gained much interest, due to its potential in improving the performance of wireless networks. The relay network, where an additional node acting as a relay is supporting the exchange of information between the network users, is an important building block for future wireless communications. \\%has attracted extensive research attention.\\ 
The capacity of the Gaussian relay network is still an elusive goal. However, Avestimehr, Diggavi and Tse presented the deterministic channel mode, through which we can get approximate results for the capacity of the Gaussian relay networks \cite{avestimehr2011wireless}. The Deterministic channel model captures the two main features of wireless communication: broadcasting and superposition of different signals. By eliminating the effect of noise, it allows us to focus only on the interactions between the different signals. Insights gleaned from the deterministic model can be used to find approximations of the capacity for the Gaussian channels.
\subsection{Related Work}
 In \cite{avestimehr2009capacity}, the authors studied the capacity region of the deterministic multi-pair bidirectional relay network, which is a generalization of the bidirectional relay channel. They proposed a simple equation-forwarding strategy that achieves this capacity region which is tight to the cut set upper bound, in which different pairs are orthogonalized on the signal level space and the relay just re-orders the received equations created from the superposition of the transmitted signals on the wireless medium and forwards them. We call this scheme the Simple Ordering Scheme (SOS). Then, from the insights of this work, the authors of \cite{hassibi2009approximate} used a combination of lattice codes and random Gaussian codes at the source nodes to propose a coding scheme that achieves to within 2 bits per user of the cut-set upper bound on the capacity of the two-pair two-way relay network. %  studied the capacity of the Gaussian two-pair full-duplex directional relay network. The authors proposed a transmission strategy using a combination of lattice codes and random Gaussian codes at the source nodes. \newline
  
 In \cite{mokhtar2010deterministic}, the authors studied the X-Channel. They considered a symmetric scenario, where the channel gain between each user and the relay is reciprocal. First, they developed a new upper bound based on the notion of a single-sided genie, then they used it to characterize the deterministic multicast capacity of their network. To prove the achievability, they proposed the idea of detour schemes that route some bits intended for a certain receiver via alternative paths when they cannot be accommodated on direct routes.\newline  
Thereafter, the capacity of the deterministic Y channel was studied in \cite{chaaban2011capacity}. Using the notion of single sided genie, the authors defined an upper bound on the capacity region, and they proved its achievability using three schemes: bi-directional, cyclic, and uni-directional communication.\newline

 Then, we extended these works in \cite{zewail2013deterministic}, by considering a 4-user relay network with no direct link, where each user wishes to exchange a number of private messages with the other 3 users via the relay node. Achievability of the capacity region was demonstrated via two Detour Schemes that are different from the ones used in \cite{mokhtar2010deterministic} due to the different nature of the multicast network. \\  
Recently, the authors studied the capacity of a 3-user relay network where the relay node is interested in exchanging some private messages with the other network users in \cite{2013itw}. It was shown that, if all messages emanating from the "relay node" are transmitted first, we obtain a reduced capacity region in the form of the one of an asymmetric 3-user relay network, which we proved its achievability of this reduced region by using a combination of the SOS and a Detour Scheme.
\subsection{Considered Scenarios and Contributions}
We consider a deterministic 4-user multicast relay network with no direct links, where each user can exchange private messages with the other network nodes. Additionally, the relay is interested in exchanging some private messages with the network users. This situation resembles the case where a base station relays messages between users and
delivers messages from the backbone system to the users as well. Furthermore, this model may represent a femtocell with intra messages and inter messages.  A distinguishing feature of this work is the assumption of non-reciprocal channels. In fact, this work is a generalization of the work in \cite{2013itw} and \cite{zewail2013deterministic}; In contrast to \cite{zewail2013deterministic}, it considers 4-users relay network with private messages and non-reciprocal channels, and in contrast to \cite{2013itw} it considers an extra user in the network. Therefore, results in \cite{mokhtar2010deterministic,chaaban2011capacity,zewail2013deterministic,2013itw} can be obtained from the results presented here with appropriate settings of system parameters.   \newline

We start by developing a new upper on the capacity region based on the notion of single sided genie. Then, we prove the achievability of this upper bound in two steps. First, relay private messages are delivered to their intended recipients. After removing the delivered messages from the network, we derive and achieve the capacity region of the resulting asymmetric 4-user relay network. This capacity is achieved by using one of the two schemes:  either the Simple Ordering Scheme (SOS) or the Detour Schemes (DS). \\

At a more fundamental level, this paper serves to show that network coding, whether through relaying messages for other users, or through aligning interference at the relay has the potential to greatly enhance network throughput. \\
Also, it's worth mentioning that we had showed the role of using Detour schemes in achieving the degrees of freedom region of the MIMO relay networks in \cite{zewailachievable}. 
%Therefore, this work can be considered as an extension of the work done in \cite{zewail2013deterministicitw} by studying a 5 node network, and as a generalization for the work done in \cite{zewail2013deterministic} by studying an asymmetric network. 
\subsection{Outline} 
In the following section, we define the system model. In Section \ref{main_result}, we state our main result, which is the capacity region of the 4-user multicast relay network with private messages from the relay. In Section \ref{achievability}, we present the first part of of our achievability proof, which results in a reduced network in the form of a 4-user non-reciprocal multicast relay network.  The capacity region of this reduced network and its achievability are studied in Section \ref{4-user}. The development of the upper bound on the capacity region based on the notion of single sided genie is explained in Section \ref{upperbound}. Section \ref{example} contains numerical examples that illustrate our achievability schemes. In Section \ref{Knode}, some comments regarding the generalization to the K user network are presented. Finally, we summarize our conclusions in Section \ref{conculsions}.   
\section{System Model}
\begin{figure}
\includegraphics[width=0.45\textwidth,height=0.16\textheight]{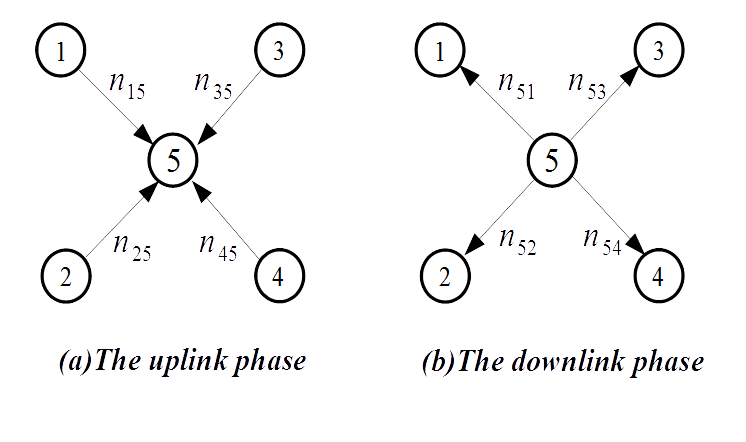}
\centering
\caption{System Model}\label{system_model}
\end{figure}

In this work, we consider a network consisting of five nodes as shown in Fig. \ref{system_model}. Each node aims to exchange private messages with the other four nodes. The nodes from 1 to 4 have no direct links between them, thus they can only communicate via node 5. However, node 5 is not merely a relay, as it has its private messages to exchange with the other four nodes. This situation resembles the case where a base station relays messages between users and delivers messages between the backbone system and the users. Also, it can be used to model a femtocell with intra messages and inter messages.\newline
Using the deterministic channel model \cite{avestimehr2011wireless}, we denote the channel gain from node $i$ to node $j$ by $n_{ij} = \lceil 0.5 \log SNR\rceil$. We assume a non-reciprocal scenario, where the $n_{i5} \ne n_{5i}$. Therefore, we assume that the uplink channel gains satisfy $n_{t5}\geq n_{u5}\geq n_{v5}\geq n_{w5}$ while the downlink channel gains satisfy $n_{5a}\geq n_{5b}\geq n_{5c}\geq n_{5d}$ where $\{t,u,v,w\},$ $\{a,b,c,d\} \in \{1,2,3,4\}$.\newline
%We assume a non-reciprocal scenario, where the channel gain between node $i$ and node 5 is not the same as the one between node $5$ and node $i$. In the deterministic model, this is equivalent to have $n_{i5}\neq n_{5i}$ where $n=\lceil 0.5 \log SNR \rceil$ and called the channel gain. Due to the assumption of non-reciprocity, the channel gains may differ between the uplink and the downlink phases. Therefore, to simplify the notation we assume without loss of generality that  and $n_{5a}\geq n_{5b}\geq n_{5c}\geq n_{5d}$ where $\{t,u,v,w\},$ $\{a,b,c,d\} \in \{1,2,3,4\}$.\newline
\section{Main Result}\label{main_result} 
%\onecolumn
\begin{thm}
The capacity region of the 4-user multicast relay network with relay private messages is given by the integral rate tuples that satisfies the inequalities (\ref{w5})-(\ref{5a10}),
%\centering
%\lipsum[1]
%\begin{widetext}

%\hline
\begin{equation}\label{w5}
R_{w5}+R_{wt}+R_{wu}+R_{wv}\leq n_{w5}
\end{equation}
\begin{equation}\label{5d}
R_{5d}+R_{ad}+R_{bd}+R_{cd}\leq n_{5d}
\end{equation}
\small
\begin{equation}\label{v5}
R_{v5}+R_{w5}+R_{wt}+R_{wu}+R_{vt}+R_{vu}+\max(R_{vw},R_{wv})\leq n_{v5}
\end{equation}
\begin{equation}\label{5c}
R_{5c}+R_{5d}+R_{ad}+R_{bd}+R_{ac}+R_{bc}+\max(R_{cd},R_{dc})\leq n_{5c}
\end{equation}
\begin{figure*}[h]
\begin{equation}\label{u51}
R_{u5}+R_{v5}+R_{w5}+R_{ut}+R_{vt}+R_{wt}+\max\left\{
                \begin{array}{ll}
                  [R_{uv}+R_{uw}+\max(R_{vw},R_{wv})],[R_{vu}+R_{vw}+\max(R_{uw},R_{wu})],\\ \qquad \qquad \qquad [R_{wu}+R_{wv}+\max(R_{uv},R_{vu})]
                \end{array}
              \right\} \leq n_{u5}
\end{equation}
%\begin{equation}\label{u52}
%R_{u5}+R_{v5}+R_{w5}+R_{ut}+R_{vt}+R_{wt}+\leq n_{u5}
%\end{equation}
%\begin{equation}\label{u53}
%R_{u5}+R_{v5}+R_{w5}+R_{ut}+R_{vt}+R_{wt}+R_{wu}+R_{wv}+\max(R_{uv},R_{vu})\leq n_{u5}
%\end{equation}
\begin{equation}\label{t51}
R_{X5}+R_{tu}+R_{tv}+R_{tw}+\max\left\{
                \begin{array}{ll}
                 [R_{uv}+R_{uw}+\max(R_{vw},R_{wv})],[R_{vu}+R_{vw}+\max(R_{uw},R_{wu})],\\
                \qquad \qquad \qquad [R_{wu}+R_{wv}+\max(R_{uv},R_{vu})])
                \end{array}
              \right\} \leq n_{t5}
\end{equation}
%\begin{equation}\label{t52}
%R_{X5}+R_{tu}+R_{tv}+R_{tw}+R_{vu}+R_{vw}+\max(R_{uw},R_{wu})\leq n_{t5}
%\end{equation}
%\begin{equation}\label{t53}
%R_{X5}+R_{tu}+R_{tv}+R_{tw}+R_{wu}+R_{wv}+\max(R_{uv},R_{vu})\leq n_{t5}
%\end{equation}
\begin{equation}\label{t54}
R_{X5}+R_{ut}+R_{uv}+R_{uw}+\max\left\{
                \begin{array}{ll}
                 [R_{tv}+R_{tw}+\max(R_{vw},R_{wv})],[R_{vt}+R_{vw}+\max(R_{tw},R_{wt})],\\
                \qquad \qquad \qquad [R_{wt}+R_{wv}+\max(R_{tv},R_{vt})]
                \end{array}
              \right\} \leq n_{t5}
\end{equation}
%\begin{equation}\label{t55}
%R_{X5}+R_{ut}+R_{uv}+R_{uw}+R_{vt}+R_{vw}+\max(R_{tw},R_{wt})\leq n_{t5}
%\end{equation}
%\begin{equation}\label{t56}
%R_{X5}+R_{ut}+R_{uv}+R_{uw}+R_{wt}+R_{wv}+\max(R_{tv},R_{vt})\leq n_{t5}
%\end{equation}
\begin{equation}\label{t57}
R_{X5}+R_{vt}+R_{vu}+R_{vw}+\max\left\{
\begin{array}{ll}
                 [R_{tu}+R_{tw}+\max(R_{uw},R_{wu})],[R_{ut}+R_{uw}+\max(R_{tw},R_{wt})],\\
                \qquad \qquad \qquad [R_{wt}+R_{wu}+\max(R_{tu},R_{ut})]
                \end{array}
              \right\} \leq n_{t5}
\end{equation}
%\begin{equation}\label{t58}
%R_{X5}+R_{vt}+R_{vu}+R_{vw}+R_{ut}+R_{uw}+\max(R_{tw},R_{wt})\leq n_{t5}
%\end{equation}
%\begin{equation}\label{t59}
%R_{X5}+R_{vt}+R_{vu}+R_{vw}+R_{wt}+R_{wu}+\max(R_{tu},R_{ut})\leq n_{t5}
%\end{equation}
\begin{equation}\label{t510}
R_{X5}+R_{wt}+R_{wu}+R_{wv}+\max\left\{
\begin{array}{ll}
                 [R_{tu}+R_{tv}+\max(R_{uv},R_{vu})],[R_{ut}+R_{uv}+\max(R_{tv},R_{vt})],\\
                \qquad \qquad \qquad [R_{vt}+R_{vu}+\max(R_{tu},R_{ut})]
                \end{array}
              \right\} \leq n_{t5}
\end{equation}
%\begin{equation}\label{t511}
%R_{X5}+R_{wt}+R_{wu}+R_{wv}+R_{ut}+R_{uv}+\max(R_{tv},R_{vt})\leq n_{t5}
%\end{equation}
%\begin{equation}\label{t512}
%R_{X5}+R_{wt}+R_{wu}+R_{wv}+R_{vt}+R_{vu}+\max(R_{tu},R_{ut})\leq n_{t5}
%\end{equation}
%\end{figure*}

%\begin{figure*}[h]
\begin{equation}\label{5b1}
R_{5b}+R_{5c}+R_{5d}+R_{ab}+R_{ac}+R_{ad}\max\left\{
                \begin{array}{ll}
                  [R_{cb}+R_{db}+\max(R_{cd},R_{dc})],[R_{bc}+R_{dc}+\max(R_{bd},R_{db})],\\
                \qquad \qquad \qquad[R_{bd}+R_{cd}+\max(R_{bc},R_{cb})]
                \end{array}
              \right\}\leq n_{5b}
\end{equation}
%\begin{equation}\label{5b2}
%R_{5b}+R_{5c}+R_{5d}+R_{ab}+R_{ac}+R_{ad}+R_{bc}+R_{dc}+\max(R_{bd},R_{db})\leq n_{5b}
%\end{equation}
%\begin{equation}\label{5b3}
%R_{5b}+R_{5c}+R_{5d}+R_{ab}+R_{ac}+R_{ad}+R_{bd}+R_{cd}+\max(R_{bc},R_{cb})\leq n_{5b}
%\end{equation}
\begin{equation}\label{5a1}
R_{5X}+R_{ba}+R_{ca}+R_{da}+\max\left\{
\begin{array}{ll}
                 [R_{cb}+R_{db}+\max(R_{cd},R_{dc})],[R_{bc}+R_{dc}+\max(R_{bd},R_{db})],\\
                \qquad \qquad \qquad [R_{bd}+R_{cd}+\max(R_{bc},R_{cb})]
                \end{array}
              \right\}\leq n_{5a}
\end{equation}
%\begin{equation}\label{5a2}
%R_{5X}+R_{ba}+R_{ca}+R_{da}+R_{bc}+R_{dc}+\max(R_{bd},R_{db})\leq n_{5a}
%\end{equation}
%\begin{equation}\label{5a3}
%R_{5X}+R_{ba}+R_{ca}+R_{da}+R_{bd}+R_{cd}+\max(R_{bc},R_{cb})\leq n_{5a}
%\end{equation}
\begin{equation}\label{5a4}
R_{5X}+R_{ab}+R_{cb}+R_{db}+\max\left\{
\begin{array}{ll}
                 [R_{ca}+R_{da}+\max(R_{cd},R_{dc})],[R_{ac}+R_{dc}+\max(R_{ad},R_{da})],\\
                \qquad \qquad \qquad [R_{ad}+R_{cd}+\max(R_{ac},R_{ca})]
                \end{array}
              \right\}\leq n_{5a}
\end{equation}
%\begin{equation}\label{5a5}
%R_{5X}+R_{ab}+R_{cb}+R_{db}+R_{ac}+R_{dc}+\max(R_{ad},R_{da})\leq n_{5a}
%\end{equation}
%\begin{equation}\label{5a6}
%R_{5X}+R_{ab}+R_{cb}+R_{db}+R_{ad}+R_{cd}+\max(R_{ac},R_{ca})\leq n_{5a}
%\end{equation}
\begin{equation}\label{5a7}
R_{5X}+R_{ac}+R_{bc}+R_{dc}+\max\left\{
\begin{array}{ll}
                 [R_{ba}+R_{da}+\max(R_{bd},R_{db})],[R_{ab}+R_{db}+\max(R_{ad},R_{da})],\\
                \qquad \qquad \qquad [R_{ad}+R_{bd}+\max(R_{ab},R_{ba})]
                \end{array}
              \right\}\leq n_{5a}
\end{equation}
%\begin{equation}\label{5a8}
%R_{5X}+R_{ac}+R_{bc}+R_{dc}+R_{ab}+R_{db}+\max(R_{ad},R_{da})\leq n_{5a}
%\end{equation}
%\begin{equation}\label{5a9}
%R_{5X}+R_{ac}+R_{bc}+R_{dc}+R_{ad}+R_{bd}+\max(R_{ab},R_{ba})\leq n_{5a}
%\end{equation}
\begin{equation}\label{5a10}
R_{5X}+R_{ad}+R_{bd}+R_{cd}+\max\left\{
\begin{array}{ll}
                 [R_{ba}+R_{ca}+\max(R_{bc},R_{cb})],[R_{ab}+R_{cb}+\max(R_{ac},R_{ca})],\\
                \qquad \qquad \qquad [R_{ac}+R_{bc}+\max(R_{ab},R_{ba})]
                \end{array}
              \right\}\leq n_{5a}
\end{equation}
%\begin{equation}\label{5a11}
%R_{5X}+R_{ad}+R_{bd}+R_{cd}+R_{ab}+R_{cb}+\max(R_{ac},R_{ca})\leq n_{5a}
%\end{equation}
%\begin{equation}\label{5a12}
%R_{5X}+R_{ad}+R_{bd}+R_{cd}+R_{ac}+R_{bc}+\max(R_{ab},R_{ba})\leq n_{5a}
%\end{equation}
%\end{widetext}
%\lipsum[1]
\end{figure*}
\normalsize
 where $R_{ij}$ is the transmission rate from node $i$ to node $j$, $R_{X5}=R_{t5}+R_{u5}+R_{v5}+R_{w5}$, and $R_{5X}=R_{5a}+R_{5b}+R_{5c}+R_{5d}$.
\end{thm} 
\section{Achievability}\label{achievability}
\begin{figure}
\includegraphics[width=0.52\textwidth,height=0.18\textheight]{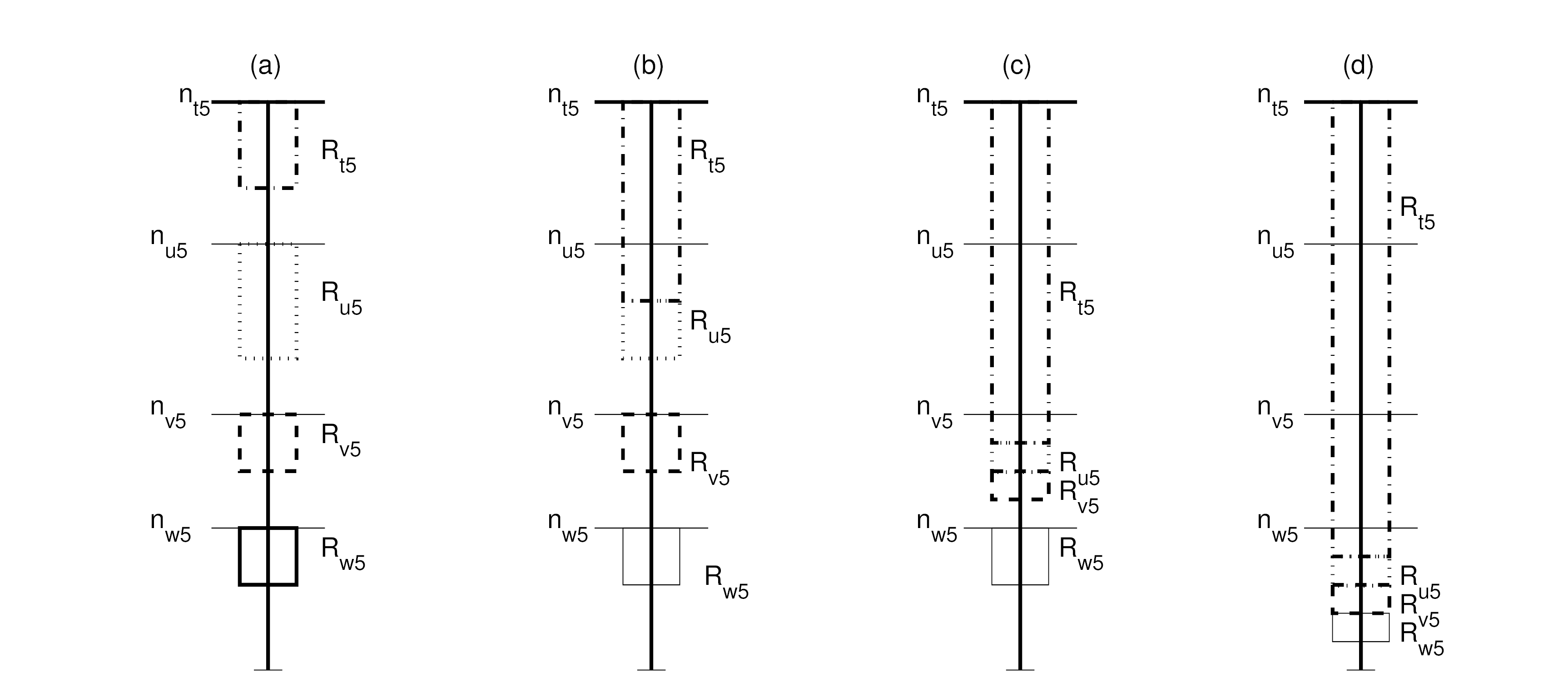}
\centering
\caption{\small Assigning levels in uplink phase: (a) $R_{t5}<n_{t5}-n_{u5}$, $R_{u5}<n_{u5}-n_{v5}$ and $R_{v5}<n_{v5}-n_{w5}$. (b) $n_{t5}-n_{v5}>R_{t5}>n_{t5}-n_{u5}$, $R_{u5}<n_{u5}-(R_{t5}-n_{t5}+n_{u5})-n_{v5}$ and $R_{v5}<n_{v5}-n_{w5}$. (c) $n_{t5}-n_{w5}>R_{t5}>n_{t5}-n_{v5}$,  $R_{u5}+R_{v5}<n_{v5}-n_{w5}-R_{t5}-(n_{t5}-n_{v5})$. (d) $R_{t5}>n_{t5}-n_{w5}$.   }\label{levelsup}
\end{figure}

\begin{figure}
\includegraphics[width=0.52\textwidth,height=0.18\textheight]{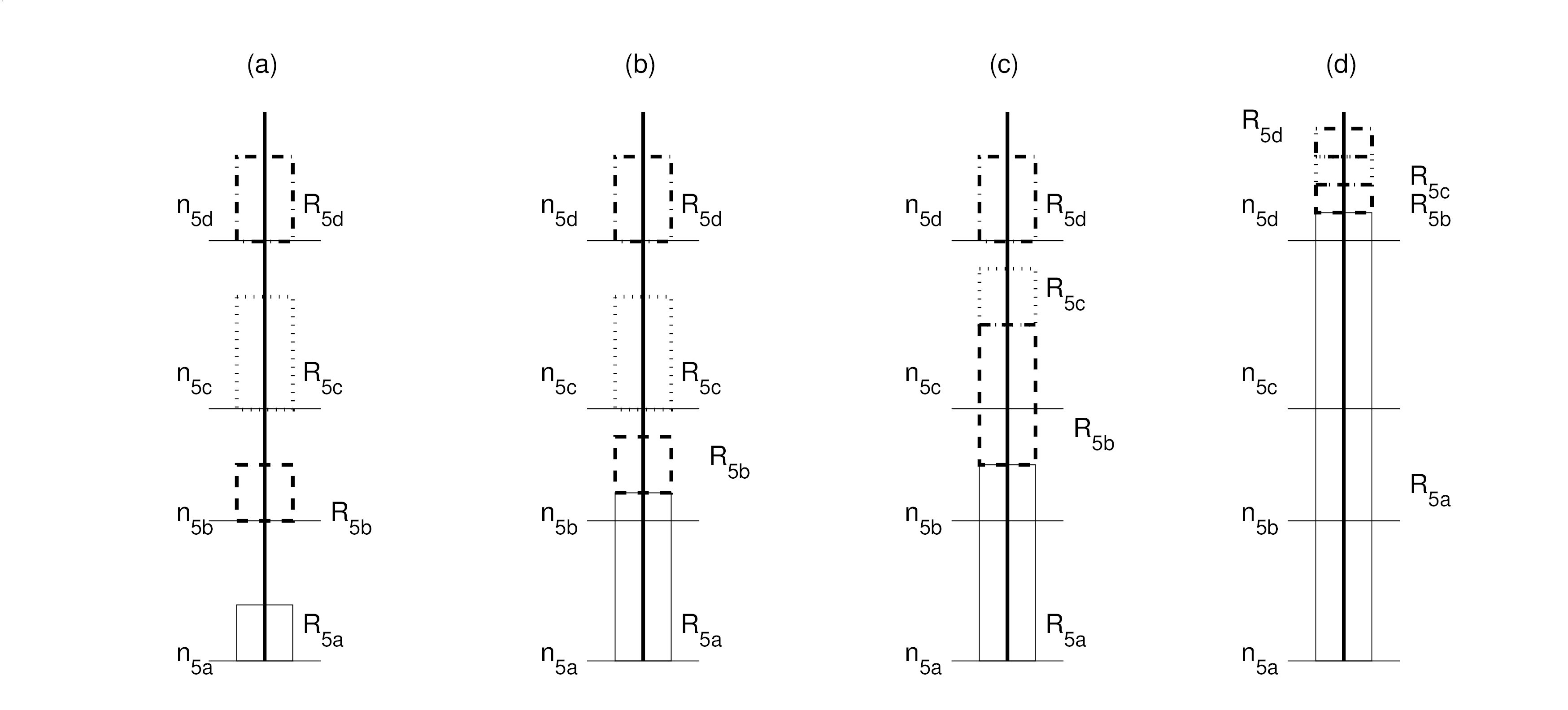}
\centering
\caption{\small Assigning levels in downlink phase: (a) $R_{5a}<n_{5a}-n_{5b}$, $R_{5b}<n_{5b}-n_{5c}$ and $R_{5c}<n_{5c}-n_{5d}$. (b) $n_{5a}-n_{5c}>R_{5a}>n_{5a}-n_{5b}$, $R_{5b}<n_{5b}-(R_{5a}-n_{5a}+n_{5b})-n_{5c}$ and $R_{5c}<n_{5c}-n_{5d}$. (c) $n_{5a}-n_{5c}>R_{5a}>n_{5a}-n_{5b}$,  $n_{5b}-n_{5d}-R_{5a}-(n_{5a}-n_{b5})<R_{5b}<n_{5b}-n_{5c}-R_{5a}-(n_{5a}-n_{b5})$, and $R_{5c}<n_{5c}-n_{5d}-R_{5b}-(n_{5b}-n_{bc})+(R_{5a}-(n_{5a}-n_{5b}))$ (d) $R_{5a}>n_{5a}-n_{5d}$.   }\label{levelsdl}
\end{figure}

In this section, we prove the achievability of all integral rate tuples that satisfy the conditions stated in Theorem 1. As a first step to our achievability scheme, we serve the uplink and downlink messages of node 5, which are represented by the rates $R_{i5}$ and $R_{5i}$, respectively. Subsequently, the network with the remaining rates is reduced to an asymmetric 4-user relay network. To complete the proof, we derive the capacity region of this reduced network in Section \ref{4-user}, and show that the reduced rate tuples are achievable. \\ %Then, in the second part, we will deal with a reduced capacity region in the form of the one represents the capacity region of an asymmetric 4-user relay network. \newline
In the uplink phase, there are a total of $n_{t5}$ levels that can be reached from all nodes. If these level are assigned indices $\ell_1$ to $\ell_{n_{t5}}$, from lowest to highest, then any node $i$ cannot send bits on levels higher than $\ell_{n_{i5}}$. Therefore, starting by node $t$, we assign levels $\ell_{n_{t5}}$ through $\ell_{n_{t5}-R_{t5}}$ to the message from node $t$ to node 5. Then, starting from level $n_{u5}$ or $n_{u5}-[R_{t5}-(n_{t5}-n_{u5})]^+-1$, whichever is smaller, we assign levels for the message $R_{5u}$ and so on.  This is illustrated by several examples in Fig. \ref{levelsup} and Fig. \ref{levelsdl}. Note that by virtue of inequalities (6) through (17), such an assignment always exists.
%In the uplink phase, we assign the highest $R_{i5}$ levels from the segment $n_{i5}$ to the messages sent from node $i$ to node 5, while in the downlink phase we assign the lowest $R_{5i}$ levels from the segment $n_{5i}$ to the messages sent from node 5 to node $i$. Note that if $R_{t5}>n_{t5}-n_{u5}$, then we serve the messages from node $u$ to node $5$ after serving the ones from node $t$ to node $5$, therefore the levels assigned for $R_{u5}$ will start from the level $n_{u5}-[R_{t5}-(n_{t5}-n_{u5})]^+-1$ from the segment $n_{u5}$, and so on for the other messages. Thus, we start to serve the messages of the node that has the highest channel gain, then we serve the node with the second highest channel gain from the first empty available level and so on for the remaining nodes in the uplink phase, and vice versa for the downlink phase.

This procedure of assigning the channel levels for node 5 messages, is equivalent to subtracting the rates related to node 5, i.e. ($R_{i5}$ and $R_{5i}$), from the both sides of all inequalities stated in Theorem 1.\newline 
By applying this subtracting operation on the conditions (\ref{t51})-(\ref{t510}) and (\ref{5a1})-(\ref{5a10}), we get the conditions (\ref{tR1})-(\ref{tR10}) and (\ref{Ra1})-(\ref{Ra10}) respectively, where $n_{tR}=n_{t5}-R_{t5}-R_{u5}-R_{v5}-R_{w5}$ and $n_{Ra}=n_{5a}-R_{5a}-R_{5b}-R_{5c}-R_{5d}$.\newline
Now, we will apply some mathematical simplifications to get a meaningful reduced region. By subtracting all rates related to node 5 from the both sides of conditions (\ref{v5}), (\ref{u51}) and (\ref{t57}), we obtain the following conditions    
%\small
\begin{equation}\label{ach1}
R_{wt}+R_{wu}+R_{vt}+R_{vu}+R_{vw}\leq n_{v5}-R_{v5}-R_{w5}
\end{equation}
\begin{equation}\nonumber%\label{ach2}
R_{wt}+R_{wu}+R_{vt}+R_{vu}+R_{vw}+R_{ut}\leq n_{u5}-R_{u5}-R_{v5}-R_{w5}
\end{equation}
\begin{equation}\nonumber%\label{ach3}
R_{wt}+R_{wu}+R_{vt}+R_{vu}+R_{vw}+R_{ut}\leq n_{t5}-R_{t5}-R_{u5}-R_{v5}-R_{w5}
\end{equation}
Since $R_{ut}\geq 0$, then we have 
\begin{equation}\label{ach2}
R_{vt}+R_{wt}+R_{vu}+R_{vw}+R_{wu}\leq n_{u5}-R_{u5}-R_{v5}-R_{w5}
\end{equation}
\begin{equation}\label{ach3}
R_{wt}+R_{wu}+R_{vt}+R_{vu}+R_{vw}\leq n_{t5}-R_{X5}
\end{equation}
%\normalsize
The conditions (\ref{ach1})-(\ref{ach3}) can be combined as 
\begin{equation}\label{ach4}
R_{wt}+R_{wu}+R_{vt}+R_{vu}+R_{vw}\leq n_{vR}
\end{equation}
where 
\small
% &=\min\{n_{v5}-R_{v5}-R_{w5},n_{u5}-n_{v5}+n_{v5}-R_{u5}-R_{v5}-R_{w5},n_{t5}-n_{v5}+n_{v5}-R_{t5}-R_{u5}-R_{v5}-R_{w5}\} \\
 %&=n_{v5}-R_{v5}-R_{w5}-\max(0,R_{u5}-(n_{u5}-n_{v5}),R_{t5}+R_{u5}-(n_{t5}-R_{u5}))\\
\begin{equation}\nonumber
\begin{aligned}
 n_{vR}&=\min\{n_{v5}-R_{v5}-R_{w5},n_{u5}-R_{u5}-R_{v5}-R_{w5}, n_{tR}\}\\ &=n_{v5}-R_{v5}-R_{w5}-\beta
\end{aligned}
\end{equation}
\begin{figure*}[h]
\[\beta =\begin{cases} [R_{t5}-(n_{t5}-n_{v5})]^++R_{u5}  &  \mbox{ $ R_{t5}\geq (n_{t5}-n_{v5})$} \\
 R_{u5}-(n_{u5}-[R_{t5}-(n_{t5}-n_{u5})]^+-n_{v5}) &  \mbox{ $  n_{t5}-n_{u5} \leq R_{t5}  < n_{t5}-n_{v5}$}\\
 [R_{u5}-(n_{u5}-n_{v5})]^+ &  \mbox{ $R_{t5} < n_{t5}-n_{u5}$}
  \end{cases}
\]
\end{figure*}
\normalsize
Again, by applying the same process on conditions (\ref{v5}), (\ref{u51}) and (\ref{t510}), we can get 
\begin{equation}\label{ach5}
R_{wt}+R_{wu}+R_{vt}+R_{vu}+R_{wv}\leq n_{vR}
\end{equation}
Combining (\ref{ach4}) and (\ref{ach5}), we get (\ref{vR}).\newline
Also, after subtracting the rates related to node 5 from the both sides of the conditions (\ref{u51}) and (\ref{t54}), we get 
%\small
\begin{equation}\nonumber%\label{ach6}
R_{ut}+R_{vt}+R_{wt}+R_{uv}+R_{uw}+R_{wv}\leq n_{u5}-R_{u5}-R_{v5}-R_{w5}
\end{equation}
\begin{equation}\nonumber%\label{ach7}
R_{ut}+R_{vt}+R_{wt}+R_{uv}+R_{uw}+R_{wv}\leq n_{t5}-R_{X5}
\end{equation}
%\normalsize
These two conditions can be combined as 
\begin{equation}\label{ach6}
R_{ut}+R_{vt}+R_{wt}+R_{uv}+R_{uw}+R_{wv}\leq n_{uR}
\end{equation}
where 
\small
%&=\min\{n_{u5}-R_{u5}-R_{v5}-R_{w5},n_{t5}-n_{u5}+n_{u5}-R_{t5}-R_{u5}-R_{v5}-R_{w5}\} \\
\begin{equation}\nonumber
\begin{aligned}
 n_{uR}&=\min\{n_{u5}-R_{u5}-R_{v5}-R_{w5},n_{t5}-R_{X5}\}\\&=n_{u5}-R_{u5}-R_{v5}-R_{w5}-[R_{t5}-(n_{t5}-n_{u5})]^+
\end{aligned}
\end{equation}
\normalsize
Again, by subtracting the rates related to node 5 from the both sides of the conditions (\ref{u51}) and (\ref{t54}), we can get 
\begin{equation}\label{ach7}
R_{ut}+R_{vt}+R_{wt}+R_{uv}+R_{uw}+R_{vw}\leq n_{uR}
\end{equation}
The conditions (\ref{ach6}) and (\ref{ach7}) can be combined as the condition (\ref{uR1}).\newline
Also, by applying the same process on the conditions (\ref{w5}), (\ref{v5}), (\ref{u51}) and (\ref{t510}), we get the following conditions 
%\small
\begin{equation}\label{ach8}
R_{wt}+R_{wu}+R_{wv}\leq n_{w5}-R_{w5}
\end{equation}
\begin{equation}\nonumber%\label{ach9}
R_{wt}+R_{wu}+R_{vt}+R_{vu}+R_{wv}\leq n_{v5}-R_{v5}-R_{w5}
\end{equation} 
\begin{equation}\nonumber%\label{ach10}
R_{ut}+R_{vt}+R_{wt}+R_{wu}+R_{wv}+R_{uv}\leq n_{u5}-R_{u5}-R_{v5}-R_{w5}
\end{equation}
\begin{equation}\nonumber%\label{ach11}
R_{wt}+R_{wu}+R_{wv}+R_{tu}+R_{tv}+R_{uv}\leq n_{t5}-R_{X5}
\end{equation}
%\normalsize
Since any $R_{ij}\geq 0$, we can get
\begin{equation}\label{ach9}
R_{wt}+R_{wu}+R_{wv}\leq n_{v5}-R_{v5}-R_{w5}
\end{equation} 
\begin{equation}\label{ach10}
R_{wt}+R_{wu}+R_{wv}\leq n_{u5}-R_{u5}-R_{v5}-R_{w5}
\end{equation}
\begin{equation}\label{ach11}
R_{wt}+R_{wu}+R_{wv}\leq n_{t5}-R_{X5}
\end{equation}
The conditions (\ref{ach8})-(\ref{ach11}) can be combined as 
\begin{equation}\label{ach9}
R_{wt}+R_{wu}+R_{wv}\leq n_{wR}
\end{equation} 
where 
\small
\begin{equation}\nonumber
\begin{aligned}
 n_{wR}&=\min\{n_{w5}-R_{w5},n_{v5}-R_{v5}-R_{w5},\\ & \qquad \qquad \qquad  n_{u5}-R_{u5}-R_{v5}-R_{w5},n_{tR}\} \\ 
 &=n_{w5}-R_{w5}-[\max(R_{t5}+R_{u5}+R_{v5}-(n_{t5}-n_{w5}),\\& \qquad R_{u5}+R_{v5}-(n_{u5}-n_{w5}),R_{t5}+R_{v5}-(n_{v5}-n_{w5}))]^+ 
 \end{aligned}
\end{equation}
\normalsize
which is the same as condition (\ref{wR}) in Theorem 2.\\
Proceeding similarly for the downlink conditions, we obtain (\ref{Rd})-(\ref{Ra10}), where  
\begin{equation}\nonumber
n_{Rb}=n_{5b}-R_{5d}-R_{5c}-R_{5b}-[R_{5a}-(n_{5a}-n_{5b})]^+
\end{equation}
\begin{equation}\nonumber
n_{Rc}=n_{5c}-R_{5d}-n_{5c}-\gamma
\end{equation}
\begin{figure*}[h]
\[\gamma =\begin{cases} [R_{5a}-(n_{5a}-n_{5c})]^++R_{5b}  &  \mbox{ $ R_{5a}\geq n_{5a}-n_{5c}$} \\
 R_{5b}-(n_{5b}-[R_{5a}-(n_{5a}-n_{5b})]^+-n_{5c}) &  \mbox{ $ n_{5a}-n_{5b} \leq R_{5a}  < n_{5a}-n_{5c}$}\\
 [R_{5b}-(n_{5b}-n_{5c})]^+ &  \mbox{ $R_{5a}<n_{5a}-n_{5b}$}
  \end{cases}
\]
\end{figure*}
\small
\begin{equation}\nonumber
\begin{aligned}
 n_{Rd}&=\min\{n_{5d}-R_{5d},n_{5c}-R_{5c}-R_{5d},\\ & \qquad \qquad  \qquad  n_{5b}-R_{5b}-R_{5c}-R_{5d},n_{Ra}\} \\ 
 &=n_{5d}-R_{5d}-[\max(R_{5a}+R_{5b}+R_{5c}-(n_{5a}-n_{5d}),\\&  \qquad R_{5b}+R_{5c}-(n_{5b}-n_{5d}),R_{5a}+R_{5c}-(n_{5c}-n_{5d}))]^+ 
 \end{aligned}
\end{equation}
\normalsize
Note that from the above expressions of the reduced channel gains i.e. ($n_{iR}$ and $n_{Ri}$), we can readily note that $n_{tR} \geq n_{uR}\geq n_{vR}\geq n_{wR}$ and $n_{Ra}\geq n_{Rb}\geq n_{Rc} \geq n_{Rd}$. %After serving messages related to node $5$, we can put $R_{t5}=R_{u5}=R_{v5}=R_{w5}=0$ and $R_{5a}=R_{5b}=R_{5c}=R_{5c}=0$.\\ %Now, we have a reduced capacity region which is stated in Theorem 3, we can observe that this region is equivalent to the capacity region of a 3-user relay network with channel gains  $n_{uR}, n_{vR}$ and $n_{tR}$ in the uplink and  $n_{Rw}, n_{Ry}$ and $n_{Rz}$ in the downlink. Therefore, to continue our achievability proof for our original network, we need to prove the achievability of the reduced region stated in Theorem 3 in the next section.
 Finally, we end up with the reduced capacity region which is stated in Theorem 2. We can observe that this region is in the form of the capacity region of asymmetric 4-user relay network with channel gains $n_{tR}, n_{uR}, n_{vR}$ and $n_{wR}$ in the uplink, and $n_{Ra}, n_{Rb}, n_{Rc}$ and $n_{Rd}$ in the downlink. Therefore, to continue our achievability proof of the original capacity region stated in Theorem 1, we need to prove the achievability of this reduced capacity region stated in Theorem 2, which is our task in the following sections.  
\section{Asymmetric 4-user Relay Networks}\label{4-user}
In this section, we study the capacity of the reduced network that resulted after serving the messages related to node 5. Thus, we derive the capacity of the asymmetric 4-user relay networks, which is given by the following theorem:  
\subsection{Main Result}
\begin{thm}
The capacity of the deterministic 4-user relay networks in given by the integral rate tuples that satisfy the following inequalities: 
\begin{equation}\label{wR}
R_{wt}+R_{wu}+R_{wv}\leq n_{wR}
\end{equation}
\begin{equation}\label{Rd}
R_{ad}+R_{bd}+R_{cd}\leq n_{Rd}
\end{equation}
\begin{equation}\label{vR}
R_{wt}+R_{wu}+R_{vt}+R_{vu}+\max(R_{vw},R_{wv})\leq n_{vR}
\end{equation}
\begin{equation}\label{Rc}
R_{ad}+R_{bd}+R_{ac}+R_{bc}+\max(R_{cd},R_{dc})\leq n_{Rc}
\end{equation}
\begin{figure*}
\begin{equation}\label{uR1}
R_{ut}+R_{vt}+R_{wt}+\max\left\{
                \begin{array}{ll}
                  [R_{uv}+R_{uw}+\max(R_{vw},R_{wv})],[R_{vu}+R_{vw}+\max(R_{uw},R_{wu})],\\ \qquad \qquad \qquad [R_{wu}+R_{wv}+\max(R_{uv},R_{vu})]
                \end{array}
              \right\} \leq n_{uR}
\end{equation}
%\begin{equation}\label{uR2}
%R_{ut}+R_{vt}+R_{wt}+R_{vu}+R_{vw}+\max(R_{uw},R_{wu})\leq n_{uR}
%\end{equation}
%\begin{equation}\label{uR3}
%R_{ut}+R_{vt}+R_{wt}+R_{wu}+R_{wv}+\max(R_{uv},R_{vu})\leq n_{uR}
%\end{equation}
\begin{equation}\label{tR1}
R_{tu}+R_{tv}+R_{tw}+\max\left\{
                \begin{array}{ll}
                  [R_{uv}+R_{uw}+\max(R_{vw},R_{wv})],[R_{vu}+R_{vw}+\max(R_{uw},R_{wu})],\\ \qquad \qquad \qquad [R_{wu}+R_{wv}+\max(R_{uv},R_{vu})]
                \end{array}
              \right\}\leq n_{tR}
\end{equation}
%\begin{equation}\label{tR2}
%R_{tu}+R_{tv}+R_{tw}+R_{vu}+R_{vw}+\max(R_{uw},R_{wu})\leq n_{tR}
%\end{equation}
%\begin{equation}\label{tR3}
%R_{tu}+R_{tv}+R_{tw}+R_{wu}+R_{wv}+\max(R_{uv},R_{vu})\leq n_{tR}
%\end{equation}
\begin{equation}\label{tR4}
R_{ut}+R_{uv}+R_{uw}+\max\left\{
                \begin{array}{ll}
                  [R_{tv}+R_{tw}+\max(R_{vw},R_{wv})],[R_{vt}+R_{vw}+\max(R_{tw},R_{wt})],\\ \qquad \qquad \qquad [R_{wt}+R_{wv}+\max(R_{tv},R_{vt}]
                \end{array}
              \right\}\leq n_{tR}
\end{equation}
%\begin{equation}\label{tR5}
%R_{ut}+R_{uv}+R_{uw}+R_{vt}+R_{vw}+\max(R_{tw},R_{wt})\leq n_{tR}
%\end{equation}
%\begin{equation}\label{tR6}
%R_{ut}+R_{uv}+R_{uw}+R_{wt}+R_{wv}+\max(R_{tv},R_{vt})\leq n_{tR}
%\end{equation}
\begin{equation}\label{tR7}
R_{vt}+R_{vu}+R_{vw}+\max\left\{
                \begin{array}{ll}
                  [R_{tu}+R_{tw}+\max(R_{uw},R_{wu})],[R_{ut}+R_{uw}+\max(R_{tw},R_{wt})],\\ \qquad \qquad \qquad [R_{wt}+R_{wu}+\max(R_{tu},R_{ut})]
                \end{array}
              \right\} \leq n_{tR}
\end{equation}
%\begin{equation}\label{tR8}
%R_{vt}+R_{vu}+R_{vw}+R_{ut}+R_{uw}+\max(R_{tw},R_{wt})\leq n_{tR}
%\end{equation}
%\begin{equation}\label{tR9}
%R_{vt}+R_{vu}+R_{vw}+R_{wt}+R_{wu}+\max(R_{tu},R_{ut})\leq n_{tR}
%\end{equation}
\begin{equation}\label{tR10}
R_{wt}+R_{wu}+R_{wv}+\max\left\{
                \begin{array}{ll}
                  [R_{tu}+R_{tv}+\max(R_{uv},R_{vu})],[R_{ut}+R_{uv}+\max(R_{tv},R_{vt})],\\ \qquad \qquad \qquad [R_{vt}+R_{vu}+\max(R_{tu},R_{ut})]
                \end{array}
              \right\}\leq n_{tR}
\end{equation}
%\begin{equation}\label{tR11}
%R_{wt}+R_{wu}+R_{wv}+R_{ut}+R_{uv}+\max(R_{tv},R_{vt})\leq n_{tR}
%\end{equation}
%\begin{equation}\label{tR12}
%R_{wt}+R_{wu}+R_{wv}+R_{vt}+R_{vu}+\max(R_{tu},R_{ut})\leq n_{tR}
%\end{equation}

\begin{equation}\label{Rb1}
R_{ab}+R_{ac}+R_{ad}+\max\left\{
                \begin{array}{ll}
                  [R_{cb}+R_{db}+\max(R_{cd},R_{dc})],[R_{bc}+R_{dc}+\max(R_{bd},R_{db})],\\ \qquad \qquad \qquad [R_{bd}+R_{cd}+\max(R_{bc},R_{cb})]
                \end{array}
              \right\} \leq n_{Rb}
\end{equation}
%\begin{equation}\label{Rb2}
%R_{ab}+R_{ac}+R_{ad}+R_{bc}+R_{dc}+\max(R_{bd},R_{db})\leq n_{Rb}
%\end{equation}
%\begin{equation}\label{Rb3}
%R_{ab}+R_{ac}+R_{ad}+R_{bd}+R_{cd}+\max(R_{bc},R_{cb})\leq n_{Rb}
%\end{equation}
\begin{equation}\label{Ra1}
R_{ba}+R_{ca}+R_{da}+\max\left\{
                \begin{array}{ll}
                  [R_{cb}+R_{db}+\max(R_{cd},R_{dc})],[R_{bc}+R_{dc}+\max(R_{bd},R_{db})],\\ \qquad \qquad \qquad [R_{bd}+R_{cd}+\max(R_{bc},R_{cb})]
                \end{array}
              \right\} \leq n_{Ra}
\end{equation}
%\begin{equation}\label{Ra2}
%R_{ba}+R_{ca}+R_{da}+R_{bc}+R_{dc}+\max(R_{bd},R_{db})\leq n_{Ra}
%\end{equation}
%\begin{equation}\label{Ra3}
%R_{ba}+R_{ca}+R_{da}+R_{bd}+R_{cd}+\max(R_{bc},R_{cb})\leq n_{Ra}
%\end{equation}
\begin{equation}\label{Ra4}
R_{ab}+R_{cb}+R_{db}+\max\left\{
                \begin{array}{ll}
                  [R_{ca}+R_{da}+\max(R_{cd},R_{dc})],[R_{ac}+R_{dc}+\max(R_{ad},R_{da})],\\ \qquad \qquad \qquad [R_{ad}+R_{cd}+\max(R_{ac},R_{ca})]
                \end{array}
              \right\} \leq n_{Ra}
\end{equation}
%\begin{equation}\label{Ra5}
%R_{ab}+R_{cb}+R_{db}+R_{ac}+R_{dc}+\max(R_{ad},R_{da})\leq n_{Ra}
%\end{equation}
%\begin{equation}\label{Ra6}
%R_{ab}+R_{cb}+R_{db}+R_{ad}+R_{cd}+\max(R_{ac},R_{ca})\leq n_{Ra}
%\end{equation}
\begin{equation}\label{Ra7}
R_{ac}+R_{bc}+R_{dc}+\max\left\{
                \begin{array}{ll}
                  [R_{ba}+R_{da}+\max(R_{bd},R_{db})],[R_{ab}+R_{db}+\max(R_{ad},R_{da})],\\ \qquad \qquad \qquad [R_{ad}+R_{bd}+\max(R_{ab},R_{ba})]
                \end{array}
              \right\} \leq n_{Ra}
\end{equation}
%\begin{equation}\label{Ra8}
%R_{ac}+R_{bc}+R_{dc}+R_{ab}+R_{db}+\max(R_{ad},R_{da})\leq n_{Ra}
%\end{equation}
%\begin{equation}\label{Ra9}
%R_{ac}+R_{bc}+R_{dc}+R_{ad}+R_{bd}+\max(R_{ab},R_{ba})\leq n_{Ra}
%\end{equation}
\begin{equation}\label{Ra10}
R_{ad}+R_{bd}+R_{cd}+\max\left\{
                \begin{array}{ll}
                  [R_{ba}+R_{ca}+\max(R_{bc},R_{cb})],[R_{ab}+R_{cb}+\max(R_{ac},R_{ca})],\\ \qquad \qquad \qquad [R_{ac}+R_{bc}+\max(R_{ab},R_{ba})]
                \end{array}
              \right\} \leq n_{Ra}
\end{equation}
\end{figure*}
%\begin{equation}\label{Ra11}
%R_{ad}+R_{bd}+R_{cd}+R_{ab}+R_{cb}+\max(R_{ac},R_{ca})\leq n_{Ra}
%\end{equation}
%\begin{equation}\label{Ra12}
%R_{ad}+R_{bd}+R_{cd}+R_{ac}+R_{bc}+\max(R_{ab},R_{ba})\leq n_{Ra}
%\end{equation}
\end{thm}
\subsection{The Achievability of The 4-user Relay Network}\label{Scheme4user}
Now, we prove the achievability of all rate tuples satisfying Theorem 2 using one of two network coding schemes: either the Simple Ordering Scheme (SOS) or the Detour Schemes (DS) which attempt to find an equivalent network with modified rates that can then be accommodated by the SOS. 
% We then develop detour scheme, where bits are routed through the network to ultimately reach the intended destination with the optimal use of the available levels. As argued next, this detour approach reduces to finding an equivalent network with modified rates that satisfy the conditions of the SOS.\\
\subsubsection{The Simple Ordering Scheme (SOS)}\label{SOS}
The SOS scheme is essentially the same as the one used in \cite{avestimehr2009capacity} and \cite{mokhtar2010deterministic}, where numerous examples details its operation. Here, we only provide an overview for completeness of our achievability presentation. 
In the Simple Ordering Scheme (SOS), each node orders its transmitted bits, such that the bi-directional messages, i.e. a message from node $i$ to node $j$ and a message from node $j$ to node $i$, will be XORed over the same channel level at the relay, then the relay reorders the received combinations and broadcasts them. In other words, if two users $i$ and $j$ wish to exchange a single bit, then they will need only use one channel level. Thus, in the uplink phase, each user will send its bit over the assigned channel level, and the relay will receive $x_{ij} \oplus x_{ji}$, then in the downlink phase, since the relay does not need to decode each bit individually, it can simply broadcast $x_{ij} \oplus x_{ji}$, since user $i$ knows $x_{ij}$, it can decode $x_{ji}$, and vice verse.% for user $j$.
\subsubsection*{SOS for the Downlink}
The messages to be transmitted are divided into four segments, each of them contains a messages of a certain user. \\
The first segment contains messages intended for user $d$ and it will be constructed as follows.  Let $\phi_k = \min(R_{dk},R_{kd})$  and $\sigma_k = \max(R_{dk},R_{kd})$ for $k \in \{a,b,c\}$.\newline
We XOR the first $\phi_k$ bits in $R_{dk}$ with the corresponding bits in $R_{kd}$. This results in a segment of size ($\phi_a + \phi_b+ \phi_c$). After inserting these XORed bits, we append with any remaining single bits to be transmitted to $d$, i.e. ($\sigma_k - \phi_k$) bits from $R_{kd}$ if $R_{kd} > R_{dk}$, otherwise all bits intended for node $d$ have already been served.\\
The second, third, and fourth segments are dedicated to messages intended for nodes $c$, $b$, and $a$ respectively and are constructed in the same manner. Note that in the higher segments, we only consider the remaining bits in each stream that were not included in lower segments.
\subsubsection*{SOS for the Uplink}
The XORed bits received in the uplink phase will be used 'as-is' in downlink phase. The relay needs only to re-order these bits to match the downlink segments described in the previous subsection.\\
%Since the relay is not required to decode all the received bits individually in the uplink phase.\\ 
%It is sufficient to provide the relay with the XORed bits resulting from signal level interactions. These XORed bits will be used 'as-is' in downlink phase. The relay needs only to re-order these bits to match the downlink segments described earlier.\\
%In principle, this scheme is the same as the ones used in \cite{avestimehr2009capacity} and \cite{mokhtar2010deterministic}. 
\begin{lemm}
 The Simple Ordering Scheme (SOS) can achieve all the integral rate tuples in the intersection between the capacity region stated in Theorem 2 and the following extra conditions:
\small
\vspace{-.2 in}
\begin{multline}\label{eqndetbR}
\max\{(R_{wu}+R_{uv}+R_{vw}),(R_{uw}+R_{wv}+R_{vu})\}\\+R_{wt}+R_{ut}+R_{vt}\leq n_{uR}
\end{multline}
\vspace{-.2 in}
\begin{multline}\label{eqndetRo}
\max\{(R_{bc}+R_{cd}+R_{db}),(R_{cb}+R_{bd}+R_{dc})\}\\+R_{ab}+R_{ac}+R_{ad}\leq n_{Rb}
\end{multline}
 %\vspace{-.45 in}
% %\small
\begin{equation}\label{eqndet31}
R_{ij}+R_{jk}+R_{ki}+\max\{(R_{li}+R_{lj}+R_{lk}),(R_{il}+R_{jl}+R_{kl})\}\leq n^*
\end{equation}
% \vspace{-.45 in}
\begin{equation}\label{eqndet41}
R_{ij}+R_{jk}+R_{kl}+R_{li}+\max(R_{jl},R_{lj})+\max(R_{ik},R_{ki})\leq n^*
\end{equation}
\normalsize
for any $\{i,j,k,l\}$ $\in$ $\{1,2,3,4\}$, where $n^*=\min(n_{tR},n_{Ra})$. 
\end{lemm}
\begin{prof}
See appendix A. 
\end{prof}
\subsubsection{The Detour Schemes}\label{Detour}
Till now, we proved the achievability of the integral rate tuples that satisfy both the conditions in Theorem 2 and those in Lemma 1 simultaneously. 
By using the Detour Schemes (DS), we will prove the achievability of any integral rate tuple that satisfies the conditions in Theorem 2 but violates one or more from the conditions stated in Lemma 1. In essence, the detour scheme converts the network into an equivalent one, where (\ref{eqndetbR})-(\ref{eqndet41}) are satisfied, thus we can apply the SOS to this equivalent network.\\ %In should be mentioned that our detour schemes are different from the ones presented by Mohktar in \cite{mokhtar2010deterministic}, due to the different nature of our network. Here, we only route the violated bits through one or two 3-node cycles, so the detoured bits will be delayed by a fixed time delay equals to one time slot.\newline
Before explaining the details of our detour schemes, we first observe that the set of extra conditions represented by (\ref{eqndetbR})-(\ref{eqndet31}) contain a 3-node cycle represented by the data flow along the leading three terms in the left hand side (LHS). % correspond to data flow along this cycle.% or a 3-node cycle consisting of nodes $b, c$ and $d$ or $o, p$ and $q$.
In contrast, conditions represented by (\ref{eqndet41}) contain two 3-node cycles. For example, if $\max (R_{jl},R_{lj})=R_{jl}$ and  $\max (R_{ik},R_{ki})=R_{ki}$, then the 3-node cycles are $i, j, k$ obtained from rates  $R_{ij}, R_{jk}$ and $R_{ki}$, and the cycle  $i, j, l$ obtained from rates  $R_{ij}, R_{jl}$ and $R_{li}$. It is worth mentioning that the notion of these cycles will be important in defining our detour schemes. \\
 %It is also important to note that each condition in (\ref{eqndet31}) and (\ref{eqndet41}) may be used to represent the above mentioned cycles in the opposite direction e.g ($k, j, i$).
%\begin{figure}
%\includegraphics[width=0.4\textwidth,height=0.2\textheight]{figures/SOScycles.png}
%\centering
%\caption{4-user relay network: A diagram that illustrates the cycles in the SOS conditions}
%\label{SOScycles}
%\end{figure}
Since any achievable rate tuple may violate more than one of the conditions expressed by (\ref{eqndetbR})-(\ref{eqndet41}), we define the \textbf{Maximum Gap Condition (MGC)} as the condition having the maximum difference between the RHS and LHS of the inequalities over all the violated conditions expressed by (\ref{eqndetbR}) and (\ref{eqndet41}). According to the form of the MGC, we will use one of the two following detour schemes:% Now we have two detour schemes, depending on the MGC:
%\hspace{-0.25 in}
\subsubsection*{Detour Scheme 1 (DS 1)}
This scheme will be used when the MGC is in the form of (\ref{eqndetbR}), (\ref{eqndetRo}) or (\ref{eqndet31}) %or (\ref{eqndet32})
for a certain $\{i,j,k,l\}$. In this case, the detour will be performed over the 3-node cycle which exists in the MGC. To simplify the notation, we assume without loss of generality that ($ R_{il}+R_{jl}+R_{kl}) \leq (R_{li}+R_{lj}+R_{lk})$, hence the MGC in the form of (\ref{eqndet31}) can be written as:
\begin{equation}\nonumber
R_{ij}+R_{jk}+R_{ki}+R_{li}+R_{lj}+R_{lk}> n^*
\end{equation}
Now, we need to reduce the rates of the left hand side by subtracting $\lambda$, such that% the modified rates satisfy:
%\begin{equation}
%(R_{ij}-\alpha)+(R_{jk}-\beta)+(R_{ki}-\gamma)+R_{li}+R_{lj}+R_{lk}\leq n_1
%\end{equation}
\begin{equation}\nonumber
(R_{ij}+R_{jk}+R_{ki})-\lambda+R_{li}+R_{lj}+R_{lk}\leq n^*
\end{equation}
The subtracted $\lambda$-bits should be transmitted to their respective destinations via alternative paths (detours). Thus all rates along this detour must be increased, while at the same time satisfying the other conditions in Theorem 2, (\ref{eqndet31}) and (\ref{eqndet41}). For example, if we decide to detour $\lambda$-bits from the $R_{ki}$ via node $j$, this means each rate of $R_{kj}$ and $R_{ji}$ should be increased by $\lambda$. Therefore, whichever the rate we choose to detour this $\lambda$-bits from, the rates over the reverse cycle should be modified as:
\begin{equation}\nonumber
R_{ji}+R_{ik}+R_{kj}\rightarrow R_{ji}+R_{ik}+R_{kj}+2\lambda
\end{equation}
\subsubsection*{Detour Scheme 2 (DS 2)}
The MGC is in the form of (\ref{eqndet41}) %or (\ref{eqndet42})
for a certain $\{i,j,k,l\}$. In this case, the detour will be performed through the two 3-node cycles represented by the MGC i.e. ($k, l, j$) obtained from rates $R_{kl}, R_{lj}$ and $R_{jk}$ and ($k, l, i$) obtained from rates $R_{kl}, R_{li}$ and $ R_{ik}$. Again, we assume without loss of generality  $ \max(R_{jl},R_{lj})+\max(R_{ik},R_{ki})=R_{ik}+R_{lj}$, hence the MGC in the form of (\ref{eqndet41}) can be written as:
\begin{equation}\nonumber
R_{ij}+R_{jk}+R_{kl}+R_{li}+R_{ik}+R_{lj}\ge n^*
\end{equation}
First, we identify the 3-nodes cycles in the MGC which are
\begin{equation}\nonumber
R_{kl} \rightarrow R_{lj} \rightarrow R_{jk} \qquad \mbox{and } \qquad   R_{kl} \rightarrow R_{li} \rightarrow R_{ik}
\end{equation}
Again, we need to reduce the LHS by subtracting an integer $\alpha$ from the rates that represent the two cycles such that the reduced rates satisfy:
\begin{equation}\label{2cyc}
R_{ij}+(R_{jk}+R_{kl}+R_{li}+R_{ik}+R_{lj})-\alpha\leq n^*
\end{equation}
%Note:$R_{kl}$ is the common between 2 cycles
The omitted $\alpha$-bits from the two cycles in (\ref{2cyc}) will be detoured over two cycles. For example, we may detour $\alpha$-bits from $R_{kl}$ via users $j$ and $l$, thus the rates over this path will be modified as follows 
%\begin{equation}\nonumber
%\begin{aligned}
%& R_{kl}\rightarrow R_{kl}-\alpha \\
%R_{kj}\rightarrow R_{kj}+\gamma
%& &
%R_{jl}\rightarrow R_{jl}+\gamma
%\\
%R_{ki}\rightarrow R_{ki}+\beta
%& &
%R_{il}\rightarrow R_{il}+\beta
%\end{aligned}
%\end{equation}
%\small
\begin{align}\nonumber
 &R_{kl}\rightarrow R_{kl}-\alpha, R_{kj}\rightarrow R_{kj}+a_1, R_{jl}\rightarrow R_{jl}+a_1,
  \nonumber \\  & \qquad R_{ki}\rightarrow R_{ki}+a_2,\quad \mbox{and} \quad R_{il}\rightarrow R_{il}+a_2\nonumber
\end{align}
%\normalsize
where $\alpha=a_1+a_2$.\\
Therefore, whichever the rates we choose to detour some bits from, the rates over the reverse cycles should be increased as follows
\begin{equation}\nonumber
R_{lk}+R_{kj}+R_{jl}+R_{ki}+R_{il}\rightarrow R_{lk}+R_{kj}+R_{jl}+R_{ki}+R_{il}+2\alpha
\end{equation}
%\begin{equation}\nonumber
%R_{kj}\rightarrow R_{kj}+\gamma
%\end{equation}
%\begin{equation}\nonumber
%R_{jl}\rightarrow R_{jl}+\gamma
%\end{equation}
%\begin{equation}\nonumber
%R_{ki}\rightarrow R_{ki}+\beta
%\end{equation}
%\begin{equation}\nonumber
%R_{il}\rightarrow R_{il}+\beta
%\end{equation}
\begin{lemm}
For all integer rate tuples for the 4-node relay network satisfying Theorem 2 and where any of the conditions in Lemma 1 is violated, it is possible to modify the rates using one of the two detour schemes to find an equivalent network, which can achieve the original rate tuple via alternative paths.
\end{lemm}
% there exists an equivalent network with modified rates that satisfy (1) and the conditions in Lemma 1. This network can be obtained by applying the DS on the original one, and then, the SOS can be used to achieve these rate tuples. 
\begin{prof}
See appendix B.
\end{prof}  
Now, we have completed the achievability proof of any integral rate tuple that satisfies the conditions stated in Theorem 1 or Theorem 2, thus the converse for these two networks will be detailed in the following section. 
\section{The Upper Bound Based on the Notion of Single Sided Genie}\label{upperbound}
%It is known that we can deal with any relay channel as the combination of two channels: Multiple Access channel i.e. (Uplink) and Broadcast channel i.e. (Downlink).\\
 In the traditional cut set bounds \cite{cover2006elements}, network nodes are divided into two sets $S$ and $S^c$, which represent the transmitting and receiving nodes, respectively. As was mentioned in \cite{mokhtar2010deterministic}, in the downlink phase, it is assumed that all nodes in $S^c$ fully cooperate with each other and share all their side information. The authors refer to this type of cooperation as the two sided genie aided bound, since it may be viewed as a genie transfers the side information of each node to other nodes on the same side of the cut. Also in the uplink phase, it makes two assumptions, the first is that a genie transfers the side information of all nodes in $S^c$ to the relay. Therefore, the relay has more information than any other node in $S^c$, i.e. it is more capable than all of them. This is clearly an upper bound, since if the relay failed to decode then all other receiving nodes will also fail to decode. The second assumption is that all nodes in $S$ share all their side information. The authors in \cite{mokhtar2010deterministic} argued that this traditional cut set bound to the relay network leads to loose bounds, therefore a tighter single sided genie aided upper bound was developed. 
%\vspace{-.05 in}
\subsection{The Downlink Upper Bound}
If we consider the cut on the downlink phase with $S=\{i,j\}$ and $S^c=\{k,l,5\}$, the two sided genie, cut bound will be:% $ $
\begin{equation}\nonumber
R_{5i}+R_{5j}+R_{ki}+R_{li}+R_{kj}+R_{lj}\leq \max(n_{5i},n_{5j})
\end{equation} 
However, let us assume that the genie transfers only all data of node $i$ to node $j$ except the data represented by $R_{ij}$. Now, node $j$ has more information than before, therefore if it fails to decode $R_{ij}$, we are sure that it will fail to decode it without this additional information. Also, if it decoded $R_{ij}$ successfully, then node $j$ has its own side information in addition to all the side information of node $i$. Thus, if node $j$ fails to decode the remaining data which is sent to it and node $i$, then we are sure that node $i$ will fail too. In the deterministic model this is equivalent to
    %therefore the data sent from node $j$ to node $i$ i.e.($R_{ji}$) is not known at node $i$ a-priori. This results in a tighter inequality as follows.
\begin{equation}\nonumber
R_{5i}+R_{5j}+R_{ki}+R_{li}+R_{kj}+R_{lj}+R_{ij}\leq \max(n_{5i},n_{5j})
\end{equation}    
%However, with the one sided genie presented in \cite{mokhtar2010deterministic}, we assume that the genie transfers only all data of node $i$ to node $j$ i.e. ($R_{ij}$), therefore the data sent from node $j$ to node $i$ i.e. ($R_{ji}$) is not known at node $i$ a-priori. This results in a tighter inequality: \vspace{-.1 in}
 % $R_{ki}+R_{li}+R_{kj}+R_{lj}+R_{ji}\leq \max(n_i,n_j)$
%\begin{equation}\nonumber
%\end{equation} 
Conversely, if the genie transfers only all the data of node $j$ to node $i$, the bound will be as follows: 
\begin{equation}\nonumber
R_{5i}+R_{5j}+R_{ki}+R_{li}+R_{kj}+R_{lj}+R_{ji}\leq \max(n_{5i},n_{5j})
\end{equation} 
These two conditions can be combined as follows:
\small
\begin{equation}\nonumber
R_{5i}+R_{5j}+R_{ki}+R_{li}+R_{kj}+R_{lj}+\max(R_{ij},R_{ji})\leq \max(n_{5i},n_{5j})
\end{equation} 
\normalsize
We can notice that this is the form of condition (\ref{5c}).\\
For $S=\{i,j,k\}$ and $S^c=\{l,5\}$, if we assume that the genie transfers all data from node $i$ to nodes $j$ and $k$ i.e.  ($R_{ij}$ and $R_{ik}$), and all data from node $j$ to node $k$ i.e. ($R_{jk}$), therefore the data sent from node $k$ to nodes $i$ and $j$ i.e. ($R_{ki}$ and $R_{kj}$) is not known at nodes $i$ and $j$, and the data sent from node $j$ to node $i$ i.e. ($R_{ji}$) is not known at node $i$. This results in a tighter inequality as follows:
\small
\begin{multline}\nonumber
R_{5i}+R_{5j}+R_{5k}+R_{li}+R_{lj}+R_{lk}\\+R_{ji}+R_{ki}+R_{kj}\leq \max(n_{5i},n_{5j},n_{5k})
\end{multline}
\normalsize 
It is clear that for the $S$, $S^c$, the order in which the genie transfers the data will affect the resulting inequality. Therefore, the previous inequality represents only one of the different genie orders namely $i \rightarrow j \rightarrow k$, which must be taken into account to characterize the upper bound. Note that this is the cut that gives, through different genie orders, the conditions (\ref{5b1}).\\% [I assume the statement is correct, please think about it]. Similar cuts involving only a single node with the relay result in inequalitiesNote that this is the cut that gives us conditions (\ref{5b1})-(\ref{5b3}).\\
It should be mentioned that for $S=\{i\}$ the cut contains only one node, the single sided genie bound coincides with the traditional two sided genie. For example $S=\{i\}$ and $S^c=\{j,k,l,5\}$, we get % two sided genie, cut bound will be:% $ $
 %$ R_{ki}+R_{li}+R_{kj}+R_{lj}\leq \max(n_i,n_j)$ \newline 
\begin{equation}\nonumber
R_{5i}+R_{ji}+R_{ki}+R_{li}\leq n_{5i}
\end{equation}   
which gives us condition (\ref{5d}) in Theorem 1. \\
 In contrast, for the 4-user relay network, we need the cut around the relay, where the one sided genie bound depends only on the different genie orders. For example, if we assume the genie order $i \rightarrow j \rightarrow k \rightarrow l$, this means the genie transfers all data of node $i$ to nodes $j, k$ and $l$, then the data of node $j$ to nodes $k$ and $l$, and finally, it transfers the data from node $k$ to node $l$, therefore the one sided genie bound in this case will be:% as follows.%$ R_{li}+R_{lj}+R_{lk}+R_{ki}+R_{kj}+R_{ji}\leq n_1 $\newline
%\hspace{-0.25 in} 
\begin{equation}\nonumber
R_{li}+R_{lj}+R_{lk}+R_{ki}+R_{kj}+R_{ji}\leq n_{Ra}
\end{equation} 
which leads to conditions (\ref{Ra1})-(\ref{Ra10}) in Theorem 2.
%To get the upper bound, we should take all possible cuts and genie orders. Finally, we can write the downlink bounds as stated in (\ref{Rq}-\ref{Ro3}) and (\ref{comman}). 
%\vspace{-0.3 in}
\subsection{The Uplink Upper Bound}
%\vspace{-0.05 in}
 For the uplink, let us take $S=\{i,j\}$ and $S^c=\{k,l,5\}$ and the relay i.e. (node 5) is only the receiving node. If we assume the one sided genie from node $i$ to node $j$, the genie transfers only $R_{ij}$ to the relay. And also the relay has all the side information of nodes $k$ and $l$, hence it should be able to decode the data transmitted to them. Therefore, if it fails then nodes $k$ and $l$ will also fail. Now, the relay is in a better position to decode compared to node $i$ and it should be able to decode $R_{ji}$. \\
  In terms of the deterministic model this is equivalent to:  
\begin{equation}\nonumber
R_{i5}+R_{j5}+R_{ik}+R_{il}+R_{jk}+R_{jl}+R_{ji}\leq \max(n_{i5},n_{j5})
\end{equation} 
Again, if the genie order is exchanged, then the bound will be: %   $  R_{ik}+R_{il}+R_{jk}+R_{jl}+R_{ij}\leq \max(n_i,n_j)$ \newline
\begin{equation}\nonumber
R_{i5}+R_{j5}+R_{ik}+R_{il}+R_{jk}+R_{jl}+R_{ij}\leq \max(n_{i5},n_{j5})
\end{equation} 
These two conditions can be combined as follows:
\small
\begin{equation}\nonumber
R_{i5}+R_{j5}+R_{ik}+R_{il}+R_{jk}+R_{jl}+\max(R_{ij},R_{ji})\leq \max(n_{i5},n_{j5})
\end{equation}
\normalsize 
which gives us condition (\ref{v5}) in Theorem 1.\\
For the cut $S=\{i,j,k\}$ and $S^c=\{l,5\}$, for the genie order $i \rightarrow j \rightarrow k$, we have the following bound 
\small
\begin{multline}\nonumber
R_{i5}+R_{j5}+R_{k5}+R_{il}+R_{jl}+R_{kl}\\+R_{ki}+R_{kj}+R_{ji}\leq \max(n_{i5},n_{j5},n_{k5})
\end{multline}
\normalsize 
which leads to conditions (\ref{u51}) in Theorem 1.\\
%We proceed in a similar manner with all possible cuts and all genie orders. Finally, we can write the uplink bounds as stated in (\ref{dR}-\ref{bR3}) and (\ref{comman}).\\
Note that, for the asymmetric 4-user relay network, again here we need the cut around the relay, where the bounds depend only on the genie order, therefore from genie order $i \rightarrow j \rightarrow k \rightarrow l$ in uplink phase we get a bound restricted to $n_{tR}$, which has the same LHS of the bound we get from the reversed genie order in downlink phase i.e. ($l \rightarrow k \rightarrow j \rightarrow i$) restricted to $n_{Ra}$. Therefore, we can combine these resultant conditions, (\ref{tR1})-(\ref{tR10}) and (\ref{Ra1})-(\ref{Ra10}) in Theorem 2, to be restricted to $n^*=\min(n_{Ra},n_{tR})$.  \newline 
By taking all cuts with all possible genie orders, we get the region stated in Theorems 1 and 2. However, some cuts do not add  new constraints on the capacity region as we will explain in the following subsection.  
\subsection{Simplifying Observations}
First, we observe that we need not take into account all cuts where $S$ contains node 5 with any other node, which significantly reduces the number of inequalities to consider. For example, let $S=\{5,1\}$ and let the genie transfer only the data from node $1$ to node $5$, i.e. $R_{15}$, this will provide the following bound:
\small 
\begin{multline}\nonumber
R_{52}+R_{53}+R_{54}+R_{12}+R_{13}+R_{14}+R_{51}\\ \leq \max(n_{51},n_{52},n_{53},n_{54})+n_{15} 
\end{multline}
\normalsize
%\vspace{-.1 in}
This condition is implicitly satisfied from the resultant bounds from the cuts $S=\{5\}$ and $S=\{1\}$
\begin{equation}\nonumber
R_{51}+R_{52}+R_{53}+R_{54}\leq \max(n_{51},n_{52},n_{53},n_{54}) 
\end{equation}
\begin{equation}\nonumber
R_{15}+R_{12}+R_{13}+R_{14}\leq n_{15} 
\end{equation}
Therefore, the cut $S=\{5,1\}$ does not result in a new constraint on the capacity region.\newline  
The second observation is that some cuts do not give a new constraint on the capacity region, therefore they were not stated in Theorems 1 and 2. For example, the bound obtained from the cut $S=\{v\}$ is implicitly included in the one obtained from the cut $S=\{v,w\}$. For $S=\{v\}$, we have
\begin{equation}\nonumber
R_{vw}+R_{vu}+R_{vt} \leq n_{vR}
\end{equation}  
 Also, from the cut $S=\{v,w\}$, we get
%\small
\begin{equation}\nonumber
R_{vu}+R_{vt}+R_{wu}+R_{wt}+\max(R_{vw},R_{wv})\leq \max(n_{vR},n_{wR})
\end{equation}
%%\normalsize
Therefore, we do not need to take into account the cut $S=\{v\}$ in characterizing the capacity region stated in Theorem 2.      
%\section{Conclusion}
%The conclusion goes here.
\section{Numerical Examples}\label{example}
In this section, we present two numerical examples that illustrate our achievability schemes.\newline 
Consider the 5-node network channel gains is given as $N_{UL}=(n_{15},n_{25},n_{35},n_{45})$, $N_{DL}=(n_{51},n_{52},n_{53},n_{54})$ and a rate tuple $R$= $(R_{12}$,  $R_{13}$, $R_{14}$, $R_{15}$, $R_{21}$, $R_{23}$, $R_{24}$, $R_{25}$, $R_{31}$, $R_{32}$, $R_{34}$, $R_{35}$, $R_{41}$, $R_{42}$, $R_{43}$, $R_{45}$, $R_{51}$, $R_{52}$, $R_{53}$, $R_{54}$), while for the reduced 4-user relay network the channel gains are given as $N_{UL}^r=(n_{1R},n_{2R},n_{3R},n_{4R})$, $N_{DL}^r=(n_{R1},n_{R2},n_{R3},n_{R4})$ and a rate tuple $R_r$ = $(R_{12}$,  $R_{13}$, $R_{14}$, $R_{21}$, $R_{23}$, $R_{24}$, $R_{31}$, $R_{32}$, $R_{34}$,$R_{41}$, $R_{42}$, $R_{43}$). 
\subsection{Example 1: The use of DS1}
Consider $N_{UL}$=(11,5,7,1), $N_{DL}$=(2,8,5,11) and $R$= (2,0,1,2,0,2,1,1,1,0,1,0,0,0,0,1,1,1,1,1). After serving user 5 messages, we get a reduced network with $N_{UL}^r=(7,4,5,0)$, $N_{DL}^r=(1,5,3,7)$ and $R_r$ = (2,0,1,0,2,1,1,0,1,0,0,0) which violates some of the conditions in Lemma 1, and the MGC is %in the form (\ref{eqndetbR})
\begin{equation}\nonumber
R_{12}+R_{23}+R_{31}+R_{14}+R_{24}+R_{34}=n^*+\lambda=7+1
\end{equation} 
We detour one bit over the cycle $1\rightarrow 2 \rightarrow 3$. In particular, we will detour one bit from the rate $R_{12}$ via user 3, which results in the modified rate tuple $\bar R_r$ = (1,1,1,0,2,1,1,1,1,0,0,0). This modified rate tuple satisfies the conditions in Theorem 2 and Lemma 1, thus we can apply the SOS.% as illustrated in Fig. \ref{ex1}. 
\begin{figure}
\includegraphics[width=0.5\textwidth,height=0.22\textheight]{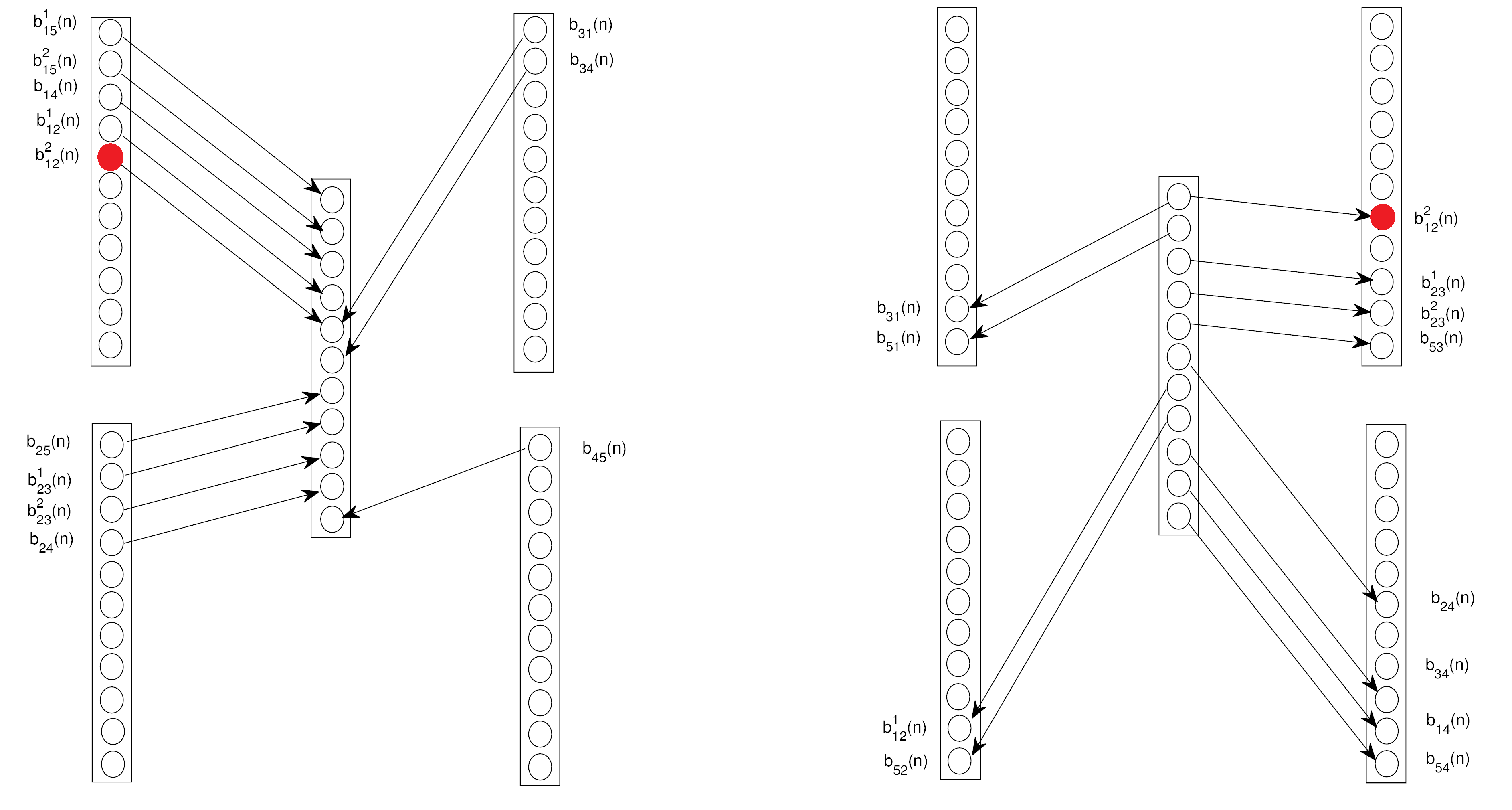}
\centering
\includegraphics[width=0.5\textwidth,height=0.22\textheight]{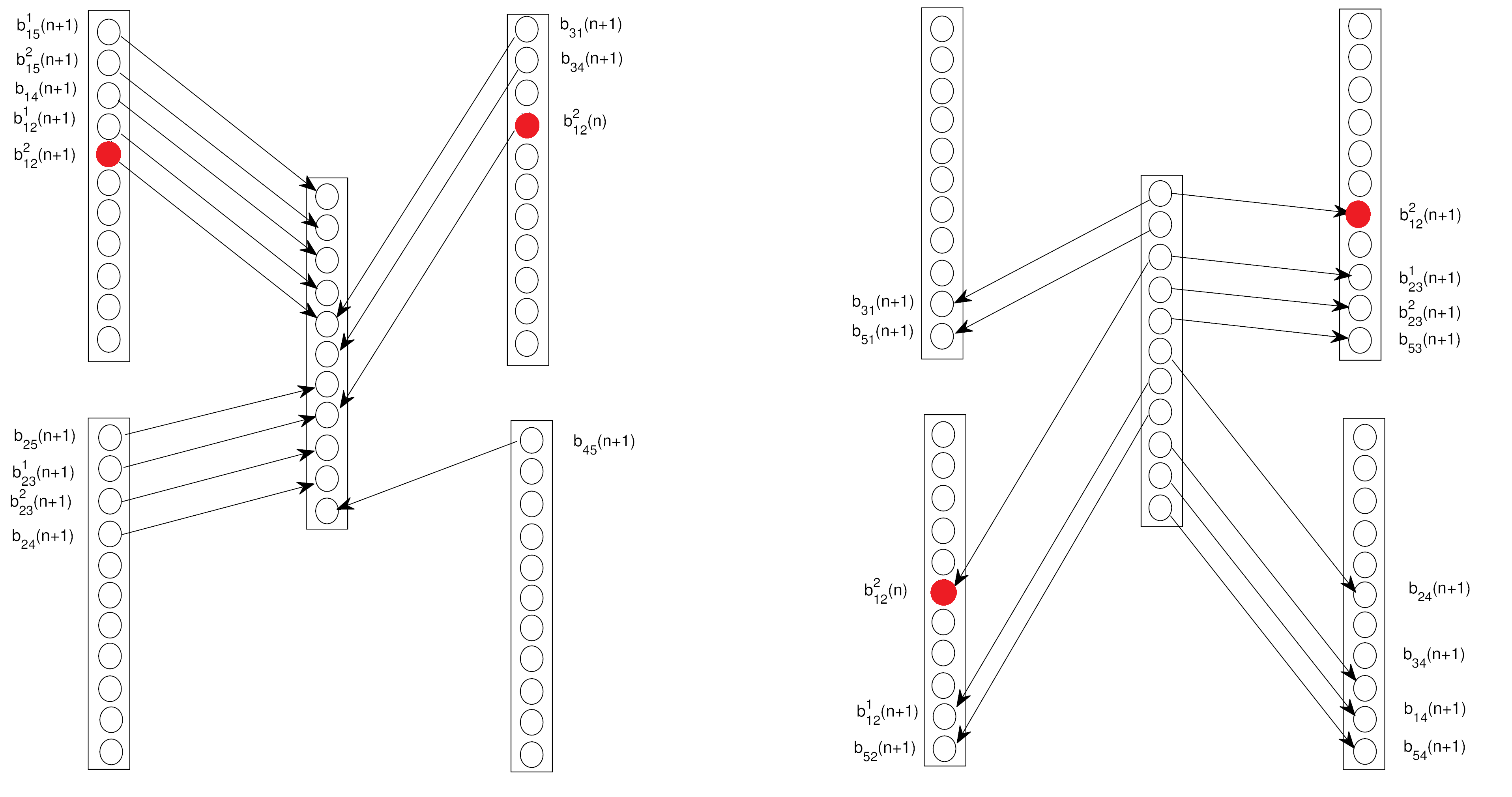}
\centering
\caption{Example on the DS1: Detouring the red bit $b_{12}^2$ via user 3}\label{ex1}
\end{figure}
\subsection{Example 2: The use of DS2}
\begin{figure}
\includegraphics[width=0.5\textwidth,height=0.22\textheight]{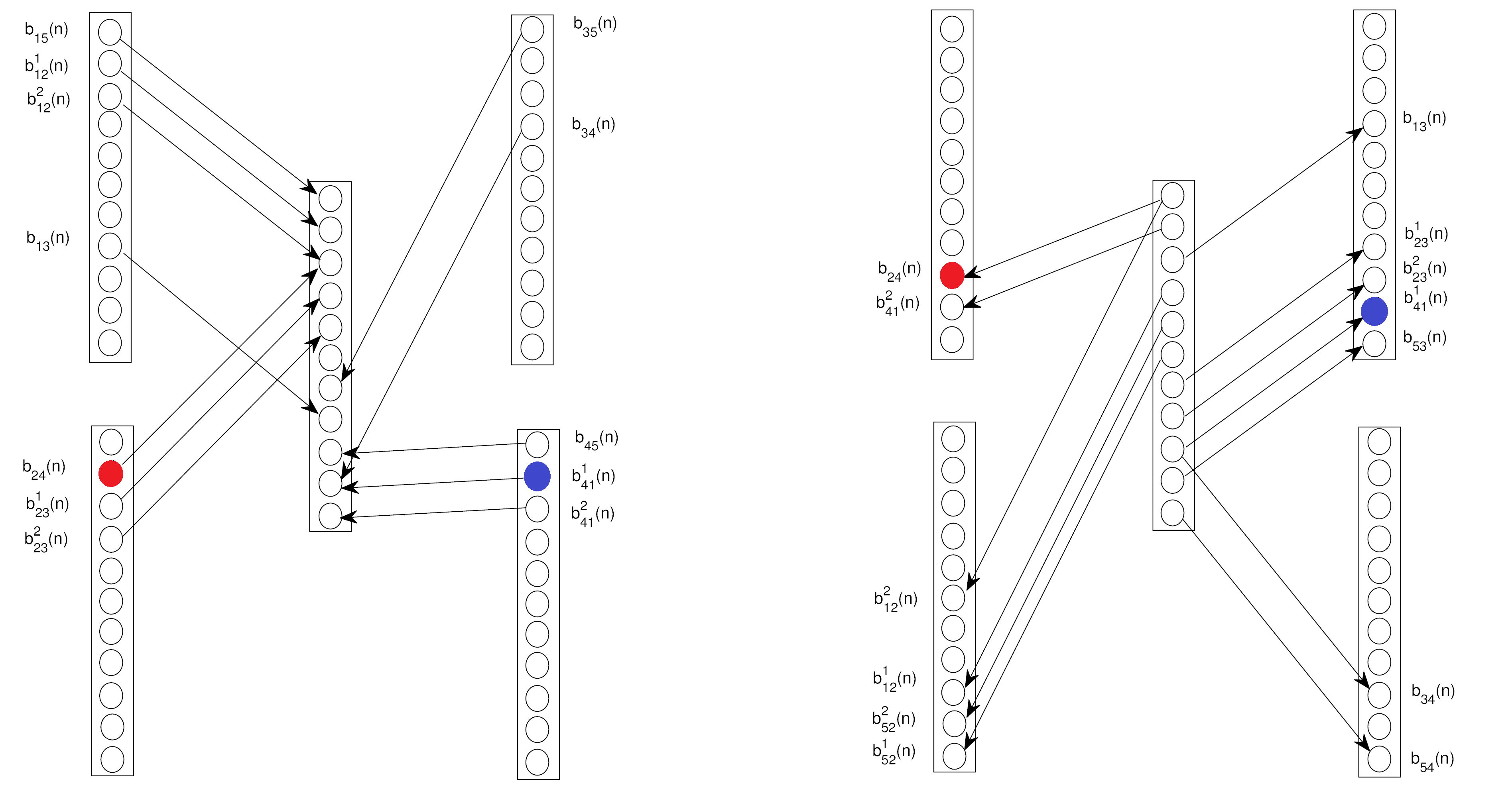}
\centering
\includegraphics[width=0.5\textwidth,height=0.22\textheight]{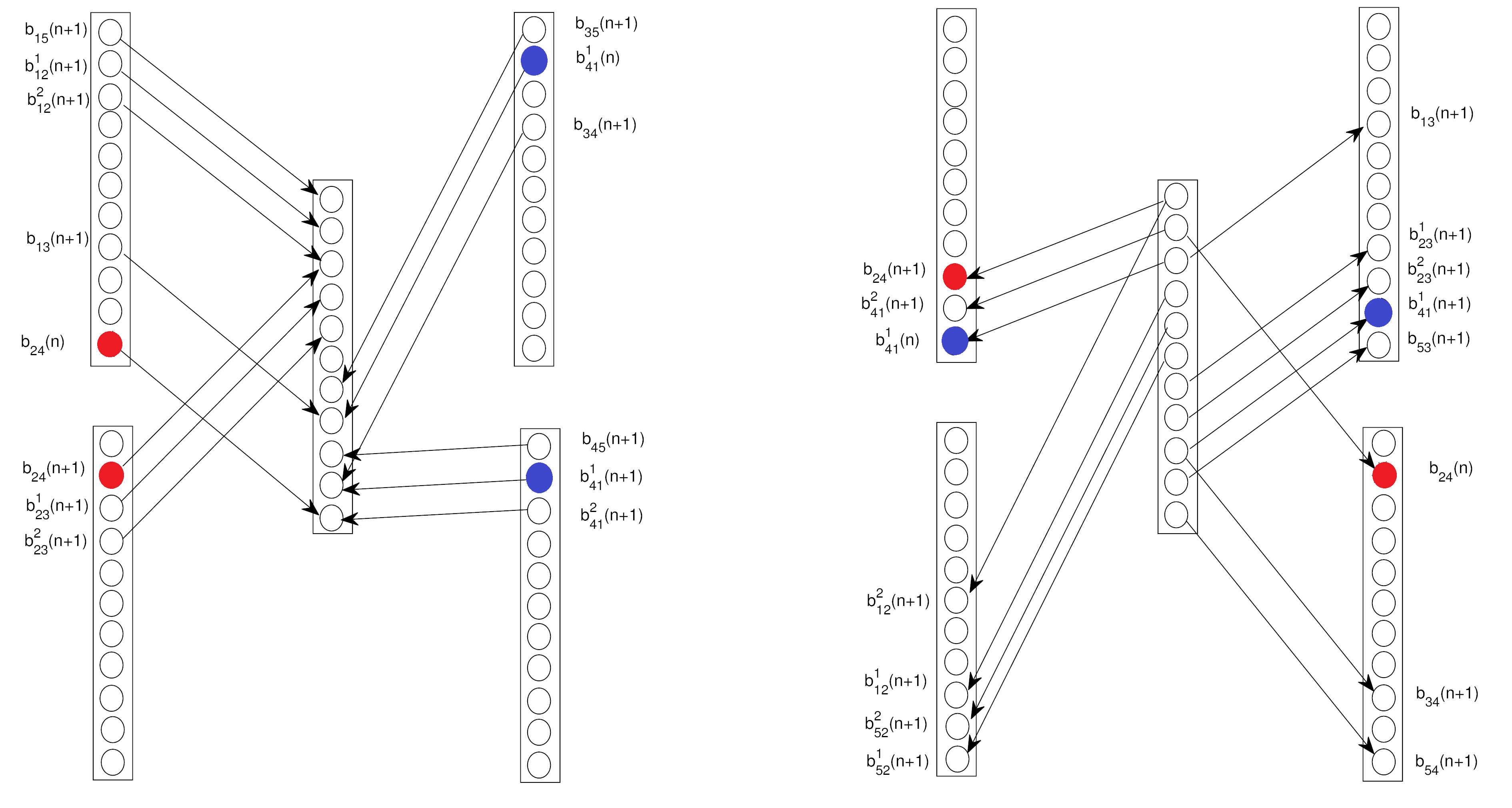}
\centering
\caption{Example on the DS2: Detouring the red bit $b_{23}^1$ via user 1 and the blue bit $b_{41}^1$ via user 3}\label{ex2}
\end{figure}

Consider $N_{UL}$=(11,10,5,3), $N_{DL}$=(3,6,10,11) and $R$= (2,1,0,1,0,2,1,0,0,0,1,1,2,0,0,1,0,2,1,1). First, we serve the messages related to node 5 in both uplink and downlink phases as illustrated in Section \ref{achievability}. Subsequently, we get a reduced network with $N_{UL}^r=(8,8,3,2)$, $N_{DL}^r=(3,4,7,7)$ and $R_r$= (2,1,0,0,2,1,0,0,1,2,0,0) which violates some of the conditions in Lemma 1, and the MGC is %in the form (\ref{eqndet41})
\begin{equation}\nonumber
R_{12}+R_{23}+R_{34}+R_{41}+R_{13}+R_{24}=n^*+\alpha=7+2
\end{equation} 
Therefore, we will detour one bit from the rate $R_{24}$ via user 1, and one bit from the rate $R_{41}$ via user 3 which results in the modified rate tuple $\bar R_r$ = (2,1,1,1,2,0,1,0,1,1,0,1). This modified rate tuple satisfies the conditions in Theorem 2 and Lemma 1, thus we can apply the SOS as illustrated in Fig. \ref{ex2}. 
% or
\section{Towards the $K$-node relay network}\label{Knode}
From the insights obtained from this paper, \cite{zewail2013deterministic} and \cite{2013itw}, we can make a conjecture about the capacity region of the $K-$node relay network with relay messages, where we have $K-1$ users $\{1,2,..,K-1\}$ each of them wishes to exchange messages with the remaining network nodes via the $K$-th node which acts as a relay beside its interest to transmit its private messages to each user. 
\subsection{The upper bound}
Based on the notion of one-sided genie, we can get the upper bound for the uplink phase as follows
%\subsubsection*{For the uplink phase} 
For the cut set that contains one node $S=\{j\}$, we get
\begin{equation}\nonumber
\sum\limits_{i=1,i\neq j}^K R_{ji} \leq n_{jK}
\end{equation}
For the cut set that contains two nodes $S=\{j,k\}$:
\small
\begin{equation}\nonumber
\sum\limits_{i=1,i\neq j}^K R_{ji}+\sum\limits_{i=1,i\neq j,k}^K R_{ki} \leq \max(n_{jK},n_{kK})
\end{equation} 
\normalsize  
For the cut set that contains three nodes $S=\{j,k,l\}$:
\small
\begin{equation}\nonumber
\sum\limits_{i=1,i\neq j}^K R_{ji}+\sum\limits_{i=1,i\neq j,k}^K R_{ki}+\sum\limits_{i=1,i\neq j,k,l}^K R_{li}\leq \max(n_{jK},n_{kK},n_{lK})
\end{equation}
\normalsize  
%Cut set contains four nodes $\{j,k,l,m\}$:
% \begin{multline}
%\sum\limits_{i=1,i\neq j}^K R_{ji}+\sum\limits_{i=1,i\neq j,k}^K R_{ki}+\sum\limits_{i=1,i\neq j,k,l}^K R_{li}+\sum\limits_{i=1,i\neq j,k,l,m}^K R_{mi}\leq \max(n_j,n_k,n_l,n_m)
% \end{multline}
%Cut set contains five nodes $\{j,k,l,m,n\}$:
% \begin{multline}
%\sum\limits_{i=1,i\neq j}^K R_{ji}+\sum\limits_{i=1,i\neq j,k}^K R_{ki}+\sum\limits_{i=1,i\neq j,k,l}^K R_{li}+\sum\limits_{i=1,i\neq j,k,l,m}^K R_{mi}\\+\sum\limits_{i=1,i\neq j,k,l,m,n}^K R_{ni} \leq \max(n_j,n_k,n_l,n_m,n_n)
% \end{multline}
And so on, for the cut set that contains $K-1$ users $S=\{1,2,..,K-1\}$:
\small
\begin{multline}\nonumber
\sum\limits_{i=1,i\neq j}^K R_{ji}+\sum\limits_{i=1,i\neq j,k}^K R_{ki}+\sum\limits_{i=1,i\neq j,k,l}^K R_{li}+\sum\limits_{i=1,i\neq j,k,l,m}^K R_{mi}\\+..+\sum\limits_{i=1,i\neq j,k,l,m,..,z}^K R_{zi}\leq \max(n_{jK},n_{kK},..,n_{zK})
\end{multline}
\normalsize
for all $\{j,k,l,..,z\} \in \{1,2,3,..,K\}$.\\ 
We can proceed similarly to derive the upper bound on the downlink phase. 
\subsection{Achievability} 
Again, after serving the $K$-th node messages, we obtain a reduced capacity region in the form of the one of an asymmetric $K-1$-user relay network. The capacity of this reduced network is obtained using a combination of the SOS and  Detour Schemes. We can conjecture that the set of extra conditions needed to apply the SOS are cyclic conditions, i.e. contain multiples of 3-node cycle. If any of these cyclic conditions is violated we will use the Detour Scheme, we will focus on forcing the MGC to be satisfied by detouring a number of bits via all 3-node cycles contained in this conditions. We tested our conjecture on 5-user relay networks, and we found that the detour can be applied over multiples of 3-node cycles, from 1 cycle up to 4 cycles, which supports our conjecture. %will be performed over multiple of 3-node cycles. 
%%%%%%%%%%%%%%%%%%%%%%%%%%%%%%%%%%%%%%%%%%%%%%%%%%%%%%%%%%%%%%%%%%%%%%%%%%%%%%
%
%Note: There are lose conditions in these conditions stated above. 
\section{Conclusions}\label{conculsions}
In this paper, we characterized the capacity region of a deterministic 4-user relay network, where the relay is interested in exchanging private messages with the network users. The use of a simplified, deterministic model, allowed derivation of exact capacity results. However, the insights gained from this simplified model suggest that cooperation among users who are not directly connected, either by relaying other users' messages (through detours) or by network coding, which took the form of XORing bits from different users in the SOS scheme, but could probably be through the use of Lattice codes in Gaussian channels, can further improve network throughput. We developed a new upper bound on the capacity region based on the notion of single sided genie. After serving the messages related to the relay node, we obtained a reduced network in the form of the asymmetric 4-user relay network. Thus, in the second part of our achievability argument, we proved the achievability of this reduced region via using the idea of the detour schemes, where we sent some bits via alternative paths instead of sending them directly. \newline 
Since we considered a 5-node network, this work strengthens the conjecture in \cite{2013itw}, that the capacity region $K$-user relay network with relay messages, can be achieved in two steps: first we serve the messages related to the relay, then we end with a capacity region of asymmetric $K$-user relay network. The capacity of this reduced network can be achieved using combination of SOS and Detour Schemes, where   the detours are performed over multiple 3-node cycles. Also, this work is a generalization for the work done in \cite{zewail2013deterministic} by considering a non-reciprocal scenario.% , since we made in We can consider this work as an extension of the work done in \cite{zewail2013deterministicitw} by studying a 5 node network, and as a generalization for the work done in \cite{zewail2013deterministic} by studying an asymmetric network.
\appendices
\section{Proof of Lemma 1}
The SOS can only work if user $i$ is able to transmit or receive all its data on the available number of levels $n_{iR}$ and $n_{Ri}$ respectively, which means that each segment can be accommodated in the corresponding channel levels. The proof depends on finding the size of each of the four segments, and applying this condition to it in both uplink and downlink phases. \\% as shown in Fig. \ref{SOSproof}. \\
%\begin{figure}.
%\begin{subfigure}{5 in }
%\includegraphics[width=0.45\textwidth,height=0.25\textheight]{figures/SOSUL4.png}
%\centering
%\end{subfigure}
%\begin{subfigure}{5 in}
%\includegraphics[width=0.45\textwidth,height=0.25\textheight]{figures/SOSDL4.png}
%\centering
%\end{subfigure}
%\caption{4-user relay network: A diagram that illustrates the main idea of the proof of Lemma 1}
%\label{SOSproof}
%\end{figure}
For the uplink phase, the condition on the size of segment $w$, which we call $SS_w$ is given by:
\begin{equation}\nonumber
SS_w= R_{wt}+R_{wu}+R_{wv}\leq n_{wR}
\end{equation}  
This condition is the same as condition (\ref{wR}) in Theorem 2.\\
Proceeding towards segment $v$, we calculate $SS_v$, then apply the condition
\begin{equation}\nonumber
SS_v+SS_w\leq n_{vR}
\end{equation}   
\begin{equation}\nonumber
\therefore R_{wt}+R_{vt}+R_{wu}+R_{vu}+\max(R_{vw},R_{wv})\leq n_{vR}
\end{equation}   
Again, this condition is the same as condition (\ref{vR}) in Theorem 2.\\
Then, we proceed towards segment $u$, we calculate $SS_u$, then apply the condition
\begin{equation}\nonumber
SS_u+SS_v+SS_w\leq n_{uR}
\end{equation} 
\small  
\begin{multline}\nonumber
\therefore R_{wt}+R_{vt}+R_{ut}+\max(R_{vw},R_{wv})\\+\max(R_{wu},R_{uw})+\max(R_{uv},R_{vu})\leq n_{uR}
\end{multline}
\normalsize 
By comparing this condition with the conditions (\ref{uR1}), we can find that we need the following extra two conditions to guarantee that the above condition is satisfied: 
\begin{equation}\nonumber
R_{wt}+R_{vt}+R_{ut}+R_{uv}+R_{vw}+R_{wu}\leq n_{uR}
\end{equation}
\begin{equation}\nonumber
R_{wt}+R_{vt}+R_{ut}+R_{vu}+R_{uw}+R_{wv}\leq n_{uR}
\end{equation}
Now, these two conditions can be combined to give condition (\ref{eqndetbR}) in Lemma 1.\\ 
Finally, we proceed towards segment $t$, we calculate $SS_t$, then apply the following condition
\begin{equation}\nonumber
SS_t+SS_u+SS_v+SS_w\leq n_{tR}
\end{equation}   
\small
\begin{multline}\nonumber
\therefore \max(R_{tu},R_{ut})+\max(R_{tv},R_{vt})+\max(R_{tw},R_{wt})\\+\max(R_{uv},R_{vu})+\max(R_{uw},R_{wu})+\max(R_{vw},R_{wv})\leq n_{tR}
\end{multline} 
\normalsize
The above condition is equivalent to a combination of $2^6=64$ conditions,  depending on the rate achieving the maximum in each of the terms above,  we are sure that 24 of them are already satisfied from the conditions (\ref{tR1})-(\ref{tR10}), thus to apply the SOS we should have rate tuple that satisfies the remaining 40 conditions, we found that these conditions can be written as (\ref{eqndet31}) and (\ref{eqndet41}) in Lemma 1, but restricted to $n_{tR}$.\\ 
We will follow the same steps for the downlink phase. Therefore, the size of segment $d$, which is denoted by $SS_d$ in downlink is given by:
\begin{equation}\nonumber
SS_d= R_{ad}+R_{bd}+R_{cd}\leq n_{Rd}
\end{equation}  
which is condition (\ref{Rd}) in Theorem 2.\\
Proceeding towards segment $c$, we calculate $SS_c$, then we apply the condition
\begin{equation}\nonumber
SS_c + SS_d\leq n_{Rc}
\end{equation}   
\begin{equation}\nonumber
\therefore R_{ac}+R_{ad}+R_{bc}+R_{bd}+\max(R_{cd},R_{dc})\leq n_{Rc}
\end{equation}   
Again, this is condition (\ref{Rc}) in Theorem 2.\\
Then, we proceed towards segment $b$, we calculate $SS_b$, then we apply the condition
\begin{equation}\nonumber
SS_b+SS_c+ SS_d\leq n_{Rb}
\end{equation} 
\small  
\begin{multline}\nonumber
\therefore R_{ab}+R_{ac}+R_{ad}+\max(R_{bc},R_{cb})\\+\max(R_{cd},R_{dc})+\max(R_{bd},R_{db})\leq n_{Rc}
\end{multline}
\normalsize  
By comparing this condition with the conditions (\ref{Rb1}), we can find that we need the following extra two conditions to apply the SOS: 
\begin{equation}\nonumber
R_{ab}+R_{ac}+R_{ad}+R_{bc}+R_{cd}+R_{db}\leq n_{Rb}
\end{equation}
\begin{equation}\nonumber
R_{ab}+R_{ac}+R_{ad}+R_{cb}+R_{bd}+R_{dc}\leq n_{Rb}
\end{equation}
These two conditions can be combined to give condition (\ref{eqndetRo}) in Lemma 1.\\ 
Finally, we proceed towards segment $a$, we calculate $SS_a$, then apply the condition
\begin{equation}\nonumber
SS_a+SS_b+SS_c+SS_d\leq n_{Ra}
\end{equation}  
\small 
\begin{multline}\nonumber
\therefore \max(R_{ab},R_{ba})+\max(R_{ac},R_{ca})+\max(R_{ad},R_{da})\\+\max(R_{bc},R_{cb})+\max(R_{bd},R_{db})+\max(R_{cd},R_{dc})\leq n_{Ra}
\end{multline}
\normalsize 
Again, by comparing the above condition with the conditions (\ref{Ra1})-(\ref{Ra10}), we can get the conditions (\ref{eqndet31}) and (\ref{eqndet41}) in Lemma 1 but restricted to $n_{Ra}$, thus if we combined them with the conditions we get from the uplink phase, we get the conditions (\ref{eqndet31}) and (\ref{eqndet41}) as stated in Lemma 1 restricted to $\min(n_{tR},n_{Ra})=n^*$.
\section{Proof of Lemma 2}
\subsection{Detour Scheme 1}
For simplicity, assume the MGC is in the form of (\ref{eqndet31}) for $\{i,j,k,l\}=\{1,2,3,4\}$ and $\max\{(R_{41}+R_{42}+R_{43}),(R_{14}+R_{24}+R_{34})\}=R_{14}+R_{24}+R_{34}$, then the MGC is% expressed as 
\begin{equation}
R_{12}+R_{23}+R_{31}+R_{14}+R_{24}+R_{34}= n^*+\lambda
\end{equation}
However, by combining the conditions (\ref{tR1})-(\ref{tR10}) with (\ref{Ra1})-(\ref{Ra10}), for any order of the channel gains, we get% the following conditions 
\begin{equation}\nonumber
R_{12}+R_{23}+R_{13}+R_{14}+R_{24}+R_{34}\leq n^*
\end{equation}
\begin{equation}\nonumber
R_{21}+R_{23}+R_{13}+R_{14}+R_{24}+R_{34}\leq n^*
\end{equation}
\begin{equation}
R_{12}+R_{32}+R_{13}+R_{14}+R_{24}+R_{34}\leq n^*
\end{equation}
By comparing these conditions with the MGC, we obtain
\begin{equation}\label{eqn: R1icondUBprof}
\begin{aligned}
R_{31} \geq R_{13}+\lambda
& & R_{12} \geq R_{21}+\lambda
& & R_{23} \geq R_{32}+\lambda
\end{aligned}
\end{equation}  
Also, from the extra SOS conditions in Lemma 1, we have
\begin{equation}\nonumber
R_{12}+R_{23}+R_{31}+R_{41}+R_{24}+R_{34}\leq n^*+\lambda
\end{equation}
\begin{equation}\nonumber
R_{12}+R_{23}+R_{31}+R_{14}+R_{42}+R_{34}\leq n^*+\lambda
\end{equation}
\begin{equation}\nonumber
R_{12}+R_{23}+R_{31}+R_{14}+R_{24}+R_{43}\leq n^*+\lambda
\end{equation}
Again, by comparing with the MGC, we get 
\begin{equation}\label{eqn: R1icondSOSprof}
\begin{aligned}
R_{14} \geq R_{41}
& & R_{24} \geq R_{42}
& & R_{34} \geq R_{43}
\end{aligned}
\end{equation}
Now, we need to choose the rate $R_{ij}$ from which we will subtract $\lambda$ bits, thus we have the following three options: 
\subsubsection{Detour bits from $R_{12}$} 
We can apply this detour as long as none of the following occurs:
\begin{itemize}
\item  $d=3$ or $w=3$.%nyNode 3 has the lowest channel gain in the uplink or the downlink phases. %$n_{3R}$ the lowest channel gain in UL
%\item $n_{R3}$ the lowest channel gain in DL
\item $c=3$ and $d=4$, or $v=3$ and $w=4$.%$n_{4R}$ the lowest channel gain in UL and $n_{3R}$ is the second lowest one. 
%\item $n_{R4}$ the lowest channel gain in DL and $n_{R3}$ is the second lowest one.
\end{itemize}
We apply the detour scheme as follows:  
\begin{equation}\nonumber
\begin{aligned}
R_{12} \rightarrow R_{12}-\lambda,
& & R_{13} \rightarrow R_{13}+\lambda
& & \text{and} & & R_{32} \rightarrow R_{32}+\lambda
\end{aligned}
\end{equation}
\subsubsection{Detour bits from $R_{23}$}
We can apply this detour as long as none of the following occurs:
\begin{itemize}
\item  $d=1$ or $w=1$.%nyNode 3 has the lowest channel gain in the uplink or the downlink phases. %$n_{3R}$ the lowest channel gain in UL
%\item $n_{R3}$ the lowest channel gain in DL
\item $c=1$ and $d=4$, or $v=1$ and $w=4$.%$n_{4R}$ the lowest channel gain in UL and $n_{3R}$ is the second lowest one. 
%\item $n_{R4}$ the lowest channel gain in DL and $n_{R3}$ is the second lowest one.
\end{itemize}
%We can apply this detour if none of the following \textbf{does not} occur:
%\begin{itemize}
%\item Node 1 has the lowest channel gain in the uplink or the downlink phases.
%\item Node 1 has the second lowest channel gain, when node 4 has the lowest one in the uplink or the downlink phases.
%%\item $n_{1R}$ the lowest channel gain in UL
%%\item $n_{R1}$ the lowest channel gain in DL
%%\item $n_{4R}$ the lowest channel gain in UL and $n_{1R}$ is the second lowest one. 
%%\item $n_{R4}$ the lowest channel gain in DL and $n_{R1}$ is the second lowest one.
%\end{itemize}  
We apply the detour scheme as follows:  
\begin{equation}\nonumber
\begin{aligned}
R_{23} \rightarrow R_{23}-\lambda,
& & R_{21} \rightarrow R_{21}+\lambda
& & \text{and} & & R_{13} \rightarrow R_{13}+\lambda
\end{aligned}
\end{equation}
\subsubsection{Detour bits from $R_{31}$}
We can apply this detour as long as none of the following occurs:
\begin{itemize}
\item  $d=2$ or $w=2$.%nyNode 3 has the lowest channel gain in the uplink or the downlink phases. %$n_{3R}$ the lowest channel gain in UL
%\item $n_{R3}$ the lowest channel gain in DL
\item $c=2$ and $d=4$, or $v=2$ and $w=4$.%$n_{4R}$ the lowest channel gain in UL and $n_{3R}$ is the second lowest one. 
%\item $n_{R4}$ the lowest channel gain in DL and $n_{R3}$ is the second lowest one.
\end{itemize}
%We can apply this detour if none of the following \textbf{does not} occur:
%\begin{itemize}
%\item Node 2 has the lowest channel gain in the uplink or the downlink phases.
%\item Node 2 has the second lowest channel gain, when node 4 has the lowest one in the uplink or the downlink phases.
%%\item $n_{2R}$ the lowest channel gain in UL
%%\item $n_{R2}$ the lowest channel gain in DL
%%\item $n_{4R}$ the lowest channel gain in UL and $n_{2R}$ is the second lowest one. 
%%\item $n_{R4}$ the lowest channel gain in DL and $n_{R2}$ is the second lowest one.
%\end{itemize}
We apply the detour scheme as follows: 
\begin{equation}\nonumber
\begin{aligned}
R_{31} \rightarrow R_{31}-\lambda,
& & R_{32} \rightarrow R_{32}+\lambda
& & \text{and} & & R_{21} \rightarrow R_{21}+\lambda
\end{aligned}
\end{equation}
By checking the conditions in Theorem 2, for each case, taking into account (\ref{eqn: R1icondUBprof}) and (\ref{eqn: R1icondSOSprof}), we can verify that all of them are satisfied. 
\textbf{Therefore, regardless the ordering of the channel gains, there will be at least one detour we can apply.}\\
\vspace{-.1 in}
\hspace{-.2 in}
\textbf{\textit{Remark.}} If the MGC is in the form of (\ref{eqndetbR}) or (\ref{eqndetRo}), the proof will follow the same steps, but we will compare the MGC with the conditions in Theorem 2 that are restricted to $n_{uR}$, i.e. (\ref{uR1}), and  $n_{Rb}$, i.e. (\ref{Rb1}), respectively.  
\vspace{-.1 in}
\subsection{Detour Scheme 2}
If the MGC is in the form of
\begin{equation}\label{eqndet41prof}
R_{ij}+R_{jk}+R_{kl}+R_{li}+\max(R_{jl},R_{lj})+\max(R_{ik},R_{ki})\leq n^*
\end{equation}
The detoured bits will be subtracted from specific rates selected from the two 3-node cycles according to the order of the channel gains in both UL and DL phases. 
For simplicity, we assume the MGC in in the form of (\ref{eqndet41prof}) for $\{i,j,k,l\}=\{1,2,3,4\}$ and $\max(R_{13},R_{31})=R_{13}$ and $\max(R_{24},R_{42})=R_{42}$, then the MGC is expressed as
\begin{equation}%\nonumber
R_{12}+R_{23}+R_{34}+R_{41}+R_{13}+R_{42}= n^*+\lambda
\end{equation}
We can observe that the MGC contains the following two 3-node cycles
\begin{equation}\nonumber
\begin{aligned}
R_{34}\rightarrow R_{42}\rightarrow R_{23}
& & &
R_{34}\rightarrow R_{41}\rightarrow R_{13}
\end{aligned}
\end{equation}
However, by combining conditions (\ref{tR1})-(\ref{tR10}) with (\ref{Ra1})-(\ref{Ra10}), for any order of the channel gains, we can obtain 
\begin{equation}
R_{43}+R_{41}+R_{42}+R_{12}+R_{13}+R_{23}\leq n^*
\end{equation} 
By comparing this condition with the MGC, we get
\begin{equation}\label{eqn: R4condUBprof}
R_{34} \geq R_{43}+\lambda
\end{equation}
From the extra SOS conditions, we have 
\begin{equation}\label{eq1}
R_{34}+R_{42}+R_{23}+R_{12}+R_{13}+R_{14}=n^*+\beta_1
\end{equation}
By subtracting this condition from the MGC, we get:
\begin{equation}\label{eqn: R4condSOS1prof}
R_{41}=R_{14}+\lambda-\beta_1
\end{equation}
Also, from the upper bound conditions in Theorem 2, we have 
\begin{equation}\nonumber
R_{12}+R_{13}+R_{14}+R_{23}+R_{24}+R_{34}\leq n^*
\end{equation}
and by comparing with (\ref{eq1}), we get 
\begin{equation}\label{eqn: R4condSOS2prof}
R_{42}\geq R_{24}+\beta_1
\end{equation}   
Again, from the extra SOS conditions in Lemma 1, we have
\begin{equation}\label{eq2}
R_{34}+R_{41}+R_{13}+R_{12}+R_{42}+R_{32}=n^*+\gamma_1
\end{equation}
By subtracting this condition from the MGC, we can get:
\begin{equation}\label{eqn: R4condSOS3prof}
R_{23}=R_{32}+\lambda-\gamma_1
\end{equation}
Also, from the upper bound conditions in Theorem 2, we can get  
\begin{equation}\nonumber
R_{31}+R_{32}+R_{34}+R_{41}+R_{42}+R_{12}\leq n^*
\end{equation}
and by comparing with (\ref{eq2}), we obtain 
\begin{equation}\label{eqn: R4condSOS4prof}
 R_{13}\geq R_{31}+\gamma_1
 \end{equation}
where $\lambda=\beta_1+\gamma_1$.\\
Now, we need to choose the rates from which we will detour the $\lambda$-bits, we have the following options:
\subsubsection{Detour $\lambda$ bits from $R_{34}$} 
We can apply this detour as long as none of the following occurs:
\begin{itemize}
\item $d=1$ or $w=1$.
\item $d=2$ or $w=2$.
\end{itemize}
%We can apply this detour if none of the following \textbf{does not} occur:
%\begin{itemize}
%\item Node 1 or 2 has the lowest channel gain in the uplink or the downlink phases. % $n_{1R}$ or $n_{3R}$ the lowest channel gain  in UL
%%\item Node 2 has the lowest channel gain and node 1 has the second lowest channel gain in the  uplink or the downlink phases. 
%%\item $n_{1R}$ or $n_{2R}$ the lowest channel gain  in UL.
%%\item $n_{R1}$ or $n_{R2}$ the lowest channel gain  in DL.
%\end{itemize}
We apply the detour scheme will be as follows: 
\begin{equation}\nonumber
\begin{aligned}
&R_{34} \rightarrow R_{34}-\lambda, R_{31} \rightarrow R_{31}+\gamma_1,
 R_{14} \rightarrow R_{14}+\gamma_1,
 \\ & \qquad R_{32} \rightarrow R_{32}+\beta_1
 \quad \text{and}\quad R_{24} \rightarrow R_{24}+\beta_1
\end{aligned}
\end{equation}
\subsubsection{Detour $\gamma_1$ bits from $R_{34}$ and  $\beta_1$ bits from $R_{42}$} 
We can apply this detour as long as none of the following occurs:
\begin{itemize}
\item $d=1$ or $w=1$.
\item $d=3$ or $w=3$.
\item $c=1$ and $d=2$, or $v=1$ and $w=2$.
\end{itemize}
%We can apply this detour if none of the following \textbf{does not} occur:
%\begin{itemize}
%\item Node 1 or 3 has the lowest channel gain in the uplink or the downlink phases. % $n_{1R}$ or $n_{3R}$ the lowest channel gain  in UL
%\item Node 2 has the lowest channel gain and node 1 has the second lowest channel gain in the uplink or the downlink phases. 
%%$n_{2R}$ is the lowest channel gain  in UL and $n_{1R}$ is the second lowest channel gain
%%\item $n_{R1}$ or $n_{R3}$ the lowest channel gain  in DL
%%\item $n_{R2}$ is the lowest channel gain  in DL and $n_{R1}$ is the second lowest channel gain
%\end{itemize}
We apply the detour scheme as follows: 
\small
\begin{equation}\nonumber
\begin{aligned}
& R_{34} \rightarrow R_{34}-\gamma_1, 
 R_{31} \rightarrow R_{31}+\gamma_1,
 R_{14} \rightarrow R_{14}+\gamma_1,\\
 & R_{42} \rightarrow R_{42}-\beta_1,
 R_{43} \rightarrow R_{43}+\beta_1 \quad \text{and}
\quad R_{32} \rightarrow R_{32}+\beta_1
\end{aligned}
\end{equation}
\normalsize
\subsubsection{Detour $\gamma_1$ bits from $R_{13}$ and  $\beta_1$ bits from $R_{42}$} 
We can apply this detour as long as none of the following occurs:
\begin{itemize}
\item $d=3$ or $w=3$.
\item $d=4$ or $w=4$.
\item $c=3$ and $d=1$, or $v=3$ and $w=1$.
\item $c=4$ and $d=2$, or $v=4$ and $w=2$.
\end{itemize}
%We can apply this detour if none of the following \textbf{does not} occur:
%\begin{itemize}
%\item Node 3 or 4 has the lowest channel gain in the uplink or the downlink phases. 
%\item Node 1 has the lowest channel gain and node 3 has the second lowest channel gain in the uplink or the downlink phases. 
%\item Node 2 has the lowest channel gain and node 4 has the second lowest channel gain in the uplink or the downlink phases. 
%%\item $n_{3R}$ or $n_{4R}$ the lowest channel gain  in UL
%%\item $n_{R3}$ or $n_{R4}$ the lowest channel gain  in DL
%%\item $n_{1R}$ is the lowest channel gain  in UL and $n_{3R}$ is the second lowest channel gain
%%\item $n_{R1}$ is the lowest channel gain  in DL and $n_{R3}$ is the second lowest channel gain
%%\item $n_{2R}$ is the lowest channel gain  in UL and $n_{4R}$ is the second lowest channel gain
%%\item $n_{R2}$ is the lowest channel gain  in DL and $n_{R4}$ is the second lowest channel gain
%\end{itemize}
We apply the detour scheme as follows: 
\small
\begin{equation}\nonumber
\begin{aligned}
&R_{13} \rightarrow R_{13}-\gamma_1,
 R_{14} \rightarrow R_{14}+\gamma_1,
 R_{43} \rightarrow R_{43}+\gamma_1,\\
 &R_{42} \rightarrow R_{42}-\beta_1
 R_{43} \rightarrow R_{43}+\beta_1 \quad \text{and}
\quad R_{32} \rightarrow R_{32}+\beta_1
\end{aligned}
\end{equation}
\normalsize
\subsubsection{Detour $\gamma_1$ bits from $R_{34}$ and  $\beta_1$ bits from $R_{23}$} 
We can apply this detour as long as none of the following occurs:
\begin{itemize}
\item $d=1$ or $w=1$.
\item $d=4$ or $w=4$.
\item $c=1$ and $d=2$, or $v=1$ and $w=2$.
%\item $c=4$ and $d=2$, or $v=4$ and $w=2$.
\end{itemize}
%We can apply this detour if none of the following \textbf{does not} occur:
%\begin{itemize}
%\item Node 1 or 4 has the lowest channel gain in the uplink or the downlink phases. 
%\item Node 2 has the lowest channel gain and node 1 has the second lowest channel gain in the uplink or the downlink phases. 
%%\item Node 2 has the lowest channel gain and node 4 has the second lowest channel gain in the uplink or the downlink phases. 
%%\item $n_{1R}$ or $n_{4R}$ the lowest channel gain  in UL
%%\item $n_{R1}$ or $n_{R4}$ the lowest channel gain  in DL
%%\item $n_{2R}$ is the lowest channel gain  in UL and $n_{1R}$ is the second lowest channel gain
%%\item $n_{R2}$ is the lowest channel gain  in DL and $n_{R1}$ is the second lowest channel gain
%\end{itemize}
We apply the detour scheme as follows:
\small 
\begin{equation}\nonumber
\begin{aligned}
&R_{34} \rightarrow R_{34}-\gamma_1,
 R_{31} \rightarrow R_{31}+\gamma_1,
 R_{14} \rightarrow R_{14}+\gamma_1,\\
 &R_{23} \rightarrow R_{23}-\beta_1,
 R_{24} \rightarrow R_{24}+\beta_1 \quad \text{and}
\quad R_{43} \rightarrow R_{43}+\beta_1
\end{aligned}
\end{equation}
\normalsize
\subsubsection{Detour $\gamma_1$ bits from $R_{41}$ and  $\beta_1$ bits from $R_{34}$} 
We can apply this detour as long as none of the following occurs:
\begin{itemize}
\item $d=2$ or $w=2$.
\item $d=3$ or $w=3$.
\item $c=2$ and $d=1$, or $v=2$ and $w=1$.
%\item $c=4$ and $d=2$, or $v=4$ and $w=2$.
\end{itemize}
%We can apply this detour if none of the following \textbf{does not} occur:
%\begin{itemize}
%\item Node 2 or 3 has the lowest channel gain in the uplink or the downlink phases. 
%\item Node 1 has the lowest channel gain and node 2 has the second lowest channel gain in the uplink or the downlink phases. 
%%\item Node 2 has the lowest channel gain and node 4 has the second lowest channel gain in the  uplink or the downlink phases. 
%%\item $n_{2R}$ or $n_{3R}$ the lowest channel gain  in UL
%%\item $n_{R2}$ or $n_{R3}$ the lowest channel gain  in DL
%%\item $n_{1R}$ is the lowest channel gain  in UL and $n_{2R}$ is the second lowest channel gain
%%\item $n_{R1}$ is the lowest channel gain  in DL and $n_{R2}$ is the second lowest channel gain
%\end{itemize}
We apply the detour scheme as follows:
\small 
\begin{equation}\nonumber
\begin{aligned}
&R_{41} \rightarrow R_{41}-\gamma_1,
 R_{43} \rightarrow R_{43}+\gamma_1,
 R_{31} \rightarrow R_{31}+\gamma_1,\\
 &R_{34} \rightarrow R_{34}-\beta_1,
 R_{32} \rightarrow R_{32}+\beta_1 \quad \text{and}
\quad R_{24} \rightarrow R_{24}+\beta_1
\end{aligned}
\end{equation}
\normalsize
%By checking the conditions in Theorem 2 for the remaining cases, taking into account (\ref{eqn: R4condUBprof}), (\ref{eqn: R4condSOS1prof}), (\ref{eqn: R4condSOS2prof}), (\ref{eqn: R4condSOS3prof}) and (\ref{eqn: R4condSOS4prof}), we can verify that all of them are satisfied.
%\begin{equation}\nonumber
%R_{12}+R_{13}+R_{14} \leq n_{1R} 
%\end{equation}
%\begin{equation}\nonumber
%R_{41}-\gamma+R_{42}+R_{43}+\gamma \leq n_{4R}   
%\end{equation}
%
%\begin{equation}\nonumber
%R_{12}+R_{14}+R_{32}+\beta+R_{34}-\beta+\max(R_{13},R_{31}+\gamma) \leq \max(n_{1R},n_{3R}) 
%\end{equation}
%\begin{equation}\nonumber
%R_{12}+R_{13}+R_{42}+R_{43}+\gamma+\max(R_{14},R_{41}-\gamma) \leq \max(n_{1R},n_{4R}) 
%\end{equation}
%
%\begin{equation}\nonumber
%R_{21}+R_{23}+R_{41}-\gamma+R_{43}+\gamma+\max(R_{24}+\beta,R_{42}) \leq \max(n_{2R},n_{4R}) 
%\end{equation}
%\begin{equation}\nonumber
%R_{31}+\gamma+R_{32}+\beta+R_{41}-\gamma+R_{42}+\max(R_{34}-\beta,R_{43}+\gamma) \leq \max(n_{3R},n_{4R})   
%\end{equation}
%and we follow the same steps for the remaining conditions, and it will be clear that they are satisfied. 
%So we will have a problem to apply this detour if none of the following occurs :
\subsubsection{Detour $\gamma_1$ bits from $R_{41}$ and  $\beta_1$ bits from $R_{42}$}
We can apply this detour as long as none of the following occurs:
\begin{itemize}
\item $d=3$ or $w=3$.
%\item $d=4$ or $w=4$.
\item $c=3$ and $d=1$, or $v=3$ and $w=1$.
\item $c=3$ and $d=2$, or $v=3$ and $w=2$.
\end{itemize}
%We can apply this detour if none of the following \textbf{does not} occur:
% \begin{itemize}
% \item Node 3 has the lowest channel gain in the uplink or the downlink phases. 
%\item Node 1 or 2 has the lowest channel gain and node 3 has the second lowest channel gain in the uplink or the downlink phases. 
%%\item Node 2 has the lowest channel gain and node 4 has the second lowest channel gain in the  uplink or the downlink phases. 
%%\item $n_{3R}$ the lowest channel gain  in UL
%%\item $n_{R3}$ the lowest channel gain  in DL
%%\item $n_{1R}$ or $n_{2R}$ is the lowest channel gain  in UL and $n_{3R}$ is the second lowest channel gain
%%\item $n_{R1}$ or $n_{R2}$ is the lowest channel gain  in DL and $n_{R3}$ is the second lowest channel gain
%\end{itemize}
We apply the detour scheme as follows: 
\small
\begin{equation}\nonumber
\begin{aligned}
&R_{41} \rightarrow R_{41}-\gamma_1,
 R_{43} \rightarrow R_{43}+\gamma_1,
 R_{31} \rightarrow R_{31}+\gamma_1,\\
 &R_{42} \rightarrow R_{42}-\beta_1,
 R_{43} \rightarrow R_{43}+\beta_1 \quad \text{and}
\quad R_{32} \rightarrow R_{32}+\beta_1
\end{aligned}
\end{equation}
\normalsize
%By checking the conditions in Theorem 2 for the remaining cases, taking into account (\ref{eqn: R4condUBprof}), (\ref{eqn: R4condSOS1prof}), (\ref{eqn: R4condSOS2prof}), (\ref{eqn: R4condSOS3prof}) and (\ref{eqn: R4condSOS4prof}), we can verify that all of them are satisfied.
%\begin{equation}\nonumber
%R_{12}+R_{13}+R_{14} \leq n_{1R} 
%\end{equation}
%\begin{equation}\nonumber
%R_{21}+R_{23}+R_{24} \leq n_{2R}  
%\end{equation}
%\begin{equation}\nonumber
%R_{41}-\gamma+R_{42}-\beta+R_{43}+\lambda \leq n_{4R}   
%\end{equation}
%\begin{equation}\nonumber
%R_{13}+R_{14}+R_{23}+R_{24}+\max(R_{12},R_{21}) \leq \max(n_{1R},n_{2R})         
%\end{equation}
%\begin{equation}\nonumber
%R_{12}+R_{13}+R_{42}-\beta+R_{43}+\lambda+\max(R_{14},R_{41}-\gamma) \leq \max(n_{1R},n_{4R}) 
%\end{equation}
%\begin{equation}\nonumber
%R_{21}+R_{23}+R_{41}-\gamma+R_{43}+\lambda+\max(R_{24},R_{42}-\beta) \leq \max(n_{2R},n_{4R}) 
%\end{equation}
%\begin{equation}\nonumber
%R_{31}+\gamma+R_{32}+\beta+R_{41}-\gamma+R_{42}-\beta+\max(R_{34},R_{43}+\lambda) \leq \max(n_{3R},n_{4R})   
%\end{equation}
%So we will have a problem to apply this detour if none of the following occurs :
%and we follow the same steps for the remaining conditions, and it will be clear that they are satisfied.
\subsubsection{Detour $\gamma_1$ bits from $R_{41}$ and  $\beta_1$ bits from $R_{23}$} 
We can apply this detour as long as none of the following occurs:
\begin{itemize}
\item $d=3$ or $w=3$.
\item $d=4$ or $w=4$.
\item $c=4$ and $d=1$, or $v=4$ and $w=1$.
\item $c=3$ and $d=2$, or $v=3$ and $w=2$.
\end{itemize}
%We can apply this detour if none of the following \textbf{does not} occur:
%\begin{itemize}
%\item Node 3 or 4 has the lowest channel gain in the uplink or the downlink phases. 
%\item Node 1 has the lowest channel gain and node 4 has the second lowest channel gain in the uplink or the downlink phases. 
%\item Node 2 has the lowest channel gain and node 3 has the second lowest channel gain in the uplink or the downlink phases. 
%%\item $n_{3R}$ or $n_{4R}$ the lowest channel gain  in UL
%%\item $n_{R3}$ or $n_{R4}$ the lowest channel gain  in DL
%%\item $n_{1R}$  is the lowest channel gain  in UL and $n_{4R}$ is the second lowest channel gain
%%\item $n_{R1}$  is the lowest channel gain  in DL and $n_{R4}$ is the second lowest channel gain
%%\item $n_{2R}$  is the lowest channel gain  in UL and $n_{3R}$ is the second lowest channel gain
%%\item $n_{R2}$  is the lowest channel gain  in DL and $n_{R3}$ is the second lowest channel gain
%\end{itemize}
We apply the detour scheme as follows: 
\small
\begin{equation}\nonumber
\begin{aligned}
&R_{41} \rightarrow R_{41}-\gamma_1,
 R_{43} \rightarrow R_{43}+\gamma_1,
 R_{31} \rightarrow R_{31}+\gamma_1,\\
&R_{23} \rightarrow R_{23}-\beta_1,
 R_{24} \rightarrow R_{24}+\beta_1 \quad \text{and}
\quad R_{43} \rightarrow R_{43}+\beta_1
\end{aligned}
\end{equation}
\normalsize
%By checking the conditions in Theorem 2 for the remaining cases, taking into account (\ref{eqn: R4condUBprof}), (\ref{eqn: R4condSOS1prof}), (\ref{eqn: R4condSOS2prof}), (\ref{eqn: R4condSOS3prof}) and (\ref{eqn: R4condSOS4prof}), we can verify that all of them are satisfied.
%\begin{equation}\nonumber
%R_{12}+R_{13}+R_{14} \leq n_{1R} 
%\end{equation}
%\begin{equation}\nonumber
%R_{21}+R_{23}-\beta+R_{24}+\beta \leq n_{2R}  
%\end{equation}
%\begin{equation}\nonumber
%R_{13}+R_{14}+R_{23}-\beta+R_{24}+\beta+\max(R_{12},R_{21}) \leq \max(n_{1R},n_{2R})         
%\end{equation}
%\begin{equation}\nonumber
%R_{12}+R_{14}+R_{32}+R_{34}+\max(R_{13},R_{31}+\gamma) \leq \max(n_{1R},n_{3R}) 
%\end{equation}
%\begin{equation}\nonumber
%R_{21}+R_{23}-\beta+R_{41}-\gamma+R_{43}+\lambda+\max(R_{24}+\beta,R_{42}) \leq \max(n_{2R},n_{4R}) 
%\end{equation}
%So we will have a problem to apply this detour if none of the following occurs :
%and we follow the same steps for the remaining conditions, and it will be clear that they are satisfied.
\subsubsection{Detour $\gamma_1$ bits from $R_{13}$ and  $\beta_1$ bits from $R_{34}$} 
We can apply this detour as long as none of the following occurs:
\begin{itemize}
\item $d=2$ or $w=2$.
\item $d=4$ or $w=4$.
\item $c=2$ and $d=1$, or $v=2$ and $w=1$.
%\item $c=4$ and $d=2$, or $v=4$ and $w=2$.
\end{itemize}
%We can apply this detour if none of the following \textbf{does not} occur:
%\begin{itemize}
%\item Node 2 or 4 has the lowest channel gain in the uplink or the downlink phases. 
%\item Node 1 has the lowest channel gain and node 2 has the second lowest channel gain in the uplink or the downlink phases. 
%%\item Node 2 has the lowest channel gain and node 4 has the second lowest channel gain in the  uplink or the downlink phases. 
%%\item $n_{2R}$ or $n_{4R}$ the lowest channel gain in UL
%%\item $n_{R2}$ or $n_{R4}$ the lowest channel gain in DL
%%\item $n_{1R}$  is the lowest channel gain in UL and $n_{2R}$ is the second lowest channel gain
%%\item $n_{R1}$  is the lowest channel gain  in DL and $n_{R2}$ is the second lowest channel gain
%\end{itemize}
We apply the detour scheme as follows:
\small 
\begin{equation}\nonumber
\begin{aligned}
&R_{13} \rightarrow R_{13}-\gamma_1,
 R_{14} \rightarrow R_{14}+\gamma_1,
 R_{43} \rightarrow R_{43}+\gamma_1,\\
 & R_{34} \rightarrow R_{34}-\beta_1,
 R_{32} \rightarrow R_{32}+\beta_1 \quad \text{and}
\quad R_{24} \rightarrow R_{24}+\beta_1
\end{aligned}
\end{equation}
\normalsize
%By checking the conditions in Theorem 2 for the remaining cases, taking into account (\ref{eqn: R4condUBprof}), (\ref{eqn: R4condSOS1prof}), (\ref{eqn: R4condSOS2prof}), (\ref{eqn: R4condSOS3prof}) and (\ref{eqn: R4condSOS4prof}), we can verify that all of them are satisfied.
%\begin{equation}\nonumber
%R_{12}+R_{13}-\gamma+R_{14}+\gamma \leq n_{1R} 
%\end{equation}
%\begin{equation}\nonumber
%R_{31}+R_{32}+\beta+R_{34}-\beta \leq n_{3R}  
%\end{equation}
%\begin{equation}\nonumber
%R_{12}+R_{14}+\gamma+R_{32}+\beta+R_{34}-\beta+\max(R_{13}-\gamma,R_{31}) \leq \max(n_{1R},n_{3R}) 
%\end{equation}
%\begin{equation}\nonumber
%R_{12}+R_{13}-\gamma+R_{42}+R_{43}+\gamma+\max(R_{14}+\gamma,R_{41}) \leq \max(n_{1R},n_{4R}) 
%\end{equation}
%\begin{equation}\nonumber
%R_{21}+R_{24}+\beta+R_{31}+R_{34}-\beta+\max(R_{23},R_{32}+\beta) \leq \max(n_{2R},n_{3R})  
%\end{equation}
%\begin{equation}\nonumber
%R_{31}+R_{32}+\beta+R_{41}+R_{42}+\max(R_{34}-\beta,R_{43}+\gamma) \leq \max(n_{3R},n_{4R})   
%\end{equation}
%So we will have a problem to apply this detour if none of the following occurs :
%and we follow the same steps for the remaining conditions, and it will be clear that they are satisfied.
\subsubsection{Detour $\gamma_1$ bits from $R_{13}$ and  $\beta_1$ bits from $R_{23}$} 
We can apply this detour as long as none of the following occurs:
\begin{itemize}
%\item $d=3$ or $w=3$.
\item $d=4$ or $w=4$.
\item $c=4$ and $d=1$, or $v=4$ and $w=1$.
\item $c=4$ and $d=2$, or $v=4$ and $w=2$.
\end{itemize}
%We can apply this detour if none of the following \textbf{does not} occur:
%\begin{itemize}
%\item Node 4 has the lowest channel gain in the uplink or the downlink phases. 
%\item Node 1 or 2 has the lowest channel gain and node 4 has the second lowest channel gain in the uplink or the downlink phases. 
%%\item Node 2 has the lowest channel gain and node 4 has the second lowest channel gain in the  uplink or the downlink phases. 
%%\item $n_{4R}$ the lowest channel gain in UL
%%\item $n_{R4}$ the lowest channel gain in DL
%%\item $n_{1R}$ or $n_{2R}$ is the lowest channel gain in UL and $n_{4R}$ is the second lowest channel gain
%%\item $n_{R1}$ or $n_{R2}$ is the lowest channel gain in DL and $n_{R4}$ is the second lowest channel gain
%\end{itemize}
We apply the detour scheme as follows: 
\small
\begin{equation}\nonumber
\begin{aligned}
&R_{13} \rightarrow R_{13}-\gamma_1 
 R_{14} \rightarrow R_{14}+\gamma_1,
 R_{43} \rightarrow R_{43}+\gamma_1,\\
 &R_{23} \rightarrow R_{23}-\beta_1,
 R_{24} \rightarrow R_{24}+\beta_1 \quad \text{and}
\quad R_{43} \rightarrow R_{43}+\beta_1
\end{aligned}
\end{equation}
\normalsize
By checking the conditions in Theorem 2, for each case, taking into account (\ref{eqn: R4condUBprof}), (\ref{eqn: R4condSOS1prof}), (\ref{eqn: R4condSOS2prof}), (\ref{eqn: R4condSOS3prof}) and (\ref{eqn: R4condSOS4prof}), we can verify that all of them are satisfied.\\
%\begin{equation}\nonumber
%R_{12}+R_{13}-\gamma+R_{14}+\gamma \leq n_{1R} 
%\end{equation}
%\begin{equation}\nonumber
%R_{21}+R_{23}-\beta+R_{24}+\beta \leq n_{2R}   
%\end{equation}
%\begin{equation}\nonumber
%R_{31}+R_{32}+R_{34} \leq n_{3R}  
%\end{equation}
%\begin{equation}\nonumber
%R_{13}-\gamma+R_{14}+\gamma+R_{23}-\beta+R_{24}+\beta+\max(R_{12},R_{21}) \leq \max(n_{1R},n_{2R}) 
%\end{equation}
%\begin{equation}\nonumber
%R_{12}+R_{14}+\gamma+R_{32}+R_{34}+\max(R_{13}-\gamma,R_{31}) \leq \max(n_{1R},n_{3R}) 
%\end{equation}
%\begin{equation}\nonumber
%R_{21}+R_{24}+\beta+R_{31}+R_{34}+\max(R_{23}-\beta,R_{32}) \leq \max(n_{2R},n_{3R})  
%\end{equation}
%\begin{equation}\nonumber
%R_{31}+R_{32}+R_{41}+R_{42}+\max(R_{34},R_{43}+\lambda) \leq \max(n_{3R},n_{4R})   
%\end{equation}
%So we will have a problem to apply this detour if none of the following occurs :
%and we follow the same steps for the remaining conditions, and it will be clear that they are satisfied.\\ 
\textbf{Therefore, regardless the ordering of the channel gains, there will be at least one detour we can apply.}
%\section{}
%\section*{Acknowledgment}
\ifCLASSOPTIONcaptionsoff
  \newpage
\fi

\bibliographystyle{IEEEtran}
\bibliography{IEEEabrv,DiversityLib}
\end{document}